\documentclass[%
 reprint,
superscriptaddress,
 amsmath,amssymb,
 aps,
floatfix,
]{revtex4-1}
\usepackage{graphicx}
\usepackage{amsmath}
\graphicspath{ {images/} }
\usepackage{dcolumn}
\usepackage{bm}
\usepackage{xcolor}
\usepackage{xr}
\usepackage{booktabs}
\usepackage{makecell}

\makeatletter
\newcommand*{\addFileDependency}[1]{
  \typeout{(#1)}
  \@addtofilelist{#1}
  \IfFileExists{#1}{}{\typeout{No file #1.}}
}
\makeatother

\begin{document}

\preprint{APS/123-QED}

\title{Novel Strontium Carbides Under Compression}
\author{Nikita Rybin}
\affiliation{Skolkovo Institute of Science and Technology, Bolshoi bulvar 30, build.1, 121205, Moscow, Russia}
\email{n.rybin@skoltech.ru}
\affiliation{Digital Materials LLC, Odintsovo, Kutuzovskaya str., 4A, 143001, Russia}

\author{Evgeny Moerman}
\affiliation{Independent researcher}
\altaffiliation{The author conducted this work at the NOMAD Laboratory at the FHI of the Max Planck Society}
\author{Pranab Gain}
\affiliation{Skolkovo Institute of Science and Technology, Bolshoi bulvar 30, build.1, 121205, Moscow, Russia}

\author{Artem R. Oganov}
\affiliation{Skolkovo Institute of Science and Technology, Bolshoi bulvar 30, build.1, 121205, Moscow, Russia}

\author{Alexander Shapeev}
\affiliation{Skolkovo Institute of Science and Technology, Bolshoi bulvar 30, build.1, 121205, Moscow, Russia}
\affiliation{Digital Materials LLC, Odintsovo, Kutuzovskaya str., 4A, 143001, Russia}
\email{a.shapeev@skoltech.ru}

\date{\today}

\begin{abstract}

Exploring the chemistry of materials at high pressures enables for the discovery of previously unknown exotic compounds. Here, we systematically search for all thermodynamically stable Sr-C compounds under pressure (up to 100 GPa) using the \textit{ab initio} evolutionary crystal structure prediction method. Our search lead to the discovery of hitherto unknown phases of SrC$_{3}$, Sr$_{2}$C$_{5}$, Sr$_{2}$C$_{3}$, Sr$_{2}$C, Sr$_{3}$C$_{2}$, and SrC. The newly discovered crystal structures feature a variety of different carbon environments ranging from isolated C anions and C-dimers to exotic polyatomic carbon anions including chains, stripes, and infinite ribbons consisting of pentagonal C$_{5}$ and hexagonal C$_{6}$ rings. Dynamical stability of all predicted compounds is confirmed by phonons calculations. Bader analysis unravels very diverse chemistry in these compounds and bonding patterns in some of them can be described using the Zintl-Klemm rule.

\end{abstract}

\pacs{Valid PACS appear here}
\maketitle

\section{Introduction}

Chemical reactions that defy conventional chemical intuition can happen under extreme conditions, resulting in the emergence of rich phase diagrams and materials possessing intriguing properties, since pressure is known to greatly affect the chemistry of elements and cause the formation of exotic compounds~\cite{mcmillan2002new, grochala2007chemical}. Well-known examples of such reactions involve the formation of sodium chlorides~\cite{zhang2013unexpected}, a stable compound of helium and sodium~\cite{dong2017stable}, platinum group metal nitrides~\cite{PhysRevLett.96.155501}, and superconducting metal hydrides~\cite{semenok2020distribution}. Nowadays, experimental high-pressure techniques allow one to explore materials chemistry up to terapascal pressure~\cite{Dubrovinsky2022} and hence, check theoretical predictions of exotic (from the chemical perspective) compounds. State-of-the-art computational techniques are reliable in predicting which compounds might be formed under certain thermodynamic conditions~\cite{oganov2019structure, miao2020chemistry}. For example, evolutionary crystal structure prediction has lead to the \textit{in silico} discovery of transparent dense sodium, unexpected forms of sodium chloride, and various superconducting metal-hydrides, which were later confirmed experimentally.~\cite{ma2009transparent, zhang2013unexpected, lilia20222021}

Among all compounds previously investigated at high pressure, a particular focus was given to carbon-containing materials. Carbon (C) itself has the ability to form sp, sp$^{2}$, and sp$^{3}$-hybridized bonds and therefore exists in various allotropic forms, including known structures of graphite, diamond, graphene, fullerenes, nanotubes.~\cite{zhu2012systematic, rice1983unusual, hoffmann1991chemical, wang2015phagraphene, li2009superhard, boulfelfel2012understanding, wang2012crystal, hoffmann1992electronic} High pressure alters the bonding patterns of carbides and leads to the formation of compounds with unusual carbon environments.~\cite{su2016catenation, krupka1969high, parasuk1989electronic} There are a large number of metal carbides containing C$_{2}$ dimers (acetylide group)~\cite{toth2014transition, spedding1958crystal, sakai1981magnetic, rundle1948structures, AKBAR2024119374, feng2018carbon}, less common is the C$_{3}$ unit (allylenide group).~\cite{west1965tetralithium, mattausch1994ho4c7, fjellvaag1992crystal, poettgen1991scandium}. For the binary systems Mg-C~\cite{strobel2014synthesis}, Ca-C~\cite{li2015investigation, Khandarkhaeva2024}, Y-C~\cite{roszak1996theoretical, Aslandukova2021, gao2014stability}, and La–C~\cite{su2016catenation}, \textit{ab initio} structure searches predicted the formation of unusual metal carbides with exotic C$_{4}$, C$_{5}$ units, C$_{6}$ rings, and graphene carbon sheets. 

Despite a thorough exploration of the chemical space of metal carbides in the past, recent experiments revealed several novel compounds in the Y-C and Ca-C systems~\cite{Aslandukova2021, Khandarkhaeva2024}. Motivated by these findings, we decided to construct a pressure-composition phase diagram of the Sr-C system. Compressed strontium shows unique structural and electronic properties~\cite{young2012phase, tsuppayakorn2015existence}, but to the best of our knowledge, strontium carbides were not investigated under compression. Here, using a variable-composition structure prediction methodology the pressure-composition phase diagram of the Sr-C system was explored in order to fully understand the structural diversity and evolution of the C-C bonding types under high pressure in the Sr-C system. This resulted in several newly predicted stable stoichiometries (SrC$_{3}$, Sr$_{2}$C$_{5}$, Sr$_{2}$C$_{3}$, Sr$_{2}$C, Sr$_{3}$C$_{2}$, and SrC) with a diverse set of carbon anions: isolated C atoms in Sr$_{2}$C, dimers in Sr$_{3}$C$_{2}$ and SrC, linear trimers in Sr$_{2}$C$_{3}$, chains in SrC, stripes with five-membered (C-pentagons) rings in SrC$_{2}$. There are also infinite ribbons consisting of five-membered (C-pentagons) rings in SrC$_{3}$, ribbons consisting of C-pentagons in SrC$_{2}$, and ribbons made of C-pentagons and C-hexagons in Sr$_{2}$C$_{5}$. Here, we discuss the structures and some properties of these compounds.

\section{Computational Methodology}

Crystal structure prediction techniques allow one to determine pressure-dependent phase diagrams for a given chemical space and a set of thermodynamic conditions. Such methodology, called a variable-composition search, is not limited to the prediction of the ground state structure of a particular stoichiometry, but explores the whole chemical space of the system in a single calculation~\cite{Oganov2011}. Recently, by exploring pressure-composition phase diagrams of Cu-F, Ag-F, and Nd-F, we have shown that a variable-composition evolutionary crystal structure search allows one to find new stable compounds even after fixed-composition searches were done~\cite{Rybin2021, Rybin2022, lilia20222021}. 

Here, we used two crystal structure prediction codes, which both implement an idea of the evolutionary crystal structure prediction. One, is the field-leading crystal structure prediction package USPEX~\cite{glass2006uspex, Oganov2006, Lyakhov2013}. Another one, is our own python-based implementation of the \textit{ab initio} evolutionary algorithm, which has a functionality to perform variable-composition search. We call this implementation Sputnik (Structure prediction using theoretical kristallography) and details of the implementation will be published elsewhere, but we have to mention that workflow management in Sputnik is done using the Fireworks functionality~\cite{Jain2015}. The theoretical basis under the Sputnik's implementation is identical to the USPEX's as presented in~\cite{PhysRevLett.75.288, Oganov2006, Oganov2011}. 

The evolutionary searches were combined with structure relaxation and energy calculations using density-functional theory (DFT) within the Perdew–Burke–Ernzerhof (PBE)~\cite{Perdew1996a} exchange–correlation functional. In the case of Sputnik, DFT calculations were performed using the plane-waves QE (Quantum Espresso) package~\cite{giannozzi2009quantum, Giannozzi_2017}, while in the case of USPEX, we used VASP (Vienna ab initio simulations package)~\cite{VASP}, with the projector augmented wave method~\cite{Blochl1994}. For the variable-composition structure search, the first generation of 160 structures was composed of randomly generated, unit cells contained up to 18 atoms (see details of the random structure generation procedure in the associated papers~\cite{Lyakhov2013, fredericks2021pyxtal}). 70$\%$ of the next generation was obtained by applying variation operators (heredity, softmutation, lattice mutation) to the 70$\%$ of the lowest-energy structures of the previous generation, while the remaining 30$\%$ of the generation were generated randomly. The percentage of structures produced by each of the variation operators was dynamically adjusted on-the-fly, based on the performance of each operator~\cite{Bushlanov2019}. During the structure search, each generation of structures was relaxed through a series of steps with increasingly more stringent calculation parameters (plane wave cutoff, k-points density, scf-cycle convergence threshold). In the last step, we did single-point calculations using a plane-wave energy cutoffs of 60~Ry (in the case of VASP the kinetic energy cutoff was set to 600~eV) and $\Gamma$-centered $k$-meshes with a reciprocal space resolution of 2$\pi\times$0.05~\r{A}$\textsuperscript{-1}$ for the Brillouin Zone sampling. In these calculations, no constraints on stoichiometries were imposed as long as the unit cell had no more than 18 atoms, and calculations were performed at 10, 25, 50, 75 and 100 GPa. 

Once structure prediction with Sputnik and USPEX was done, we calculated the pressure-composition phase diagram with a resolution of 10 GPa, then we confirmed the dynamical stability of all novel structures with phonon calculations using the supercell approach and the finite displacement method~\cite{Parlinski1997}, as implemented in the Phonopy package~\cite{Togo2015}. The electron localization function (ELF) was calculated for all the novel phases using VASP and visualized using VESTA~\cite{momma2011vesta}. Bader charge analyses was performed on total charge density~\cite{henkelman2006fast,yu2011accurate,sanville2007improved,tang2009grid}. Structural information for all discovered compounds, band structures, and phonon dispersion curves are presented in the Supplementary Materials (SM).

\section{Results}

To determine stable phases in a variable-composition system, it is convenient to use the convex hull construction. Phases located on the thermodynamic convex hull are stable with respect to decomposition into elemental Sr and C or other Sr-C compounds. All the values of the chemical potentials, delimiting the fields of stability of compounds, are derived directly from the convex hull. Consequently, the convex hull construction is the cornerstone of the variable-composition structure prediction, which determines the stability of compounds at particular thermodynamic conditions~\cite{Oganov2011}. 

\begin{figure}[h!]
	\centering
	\begin{minipage}[h]{1\linewidth}
		\center{\includegraphics[trim={0cm 0cm 0cm 0cm},clip, width=1\linewidth]{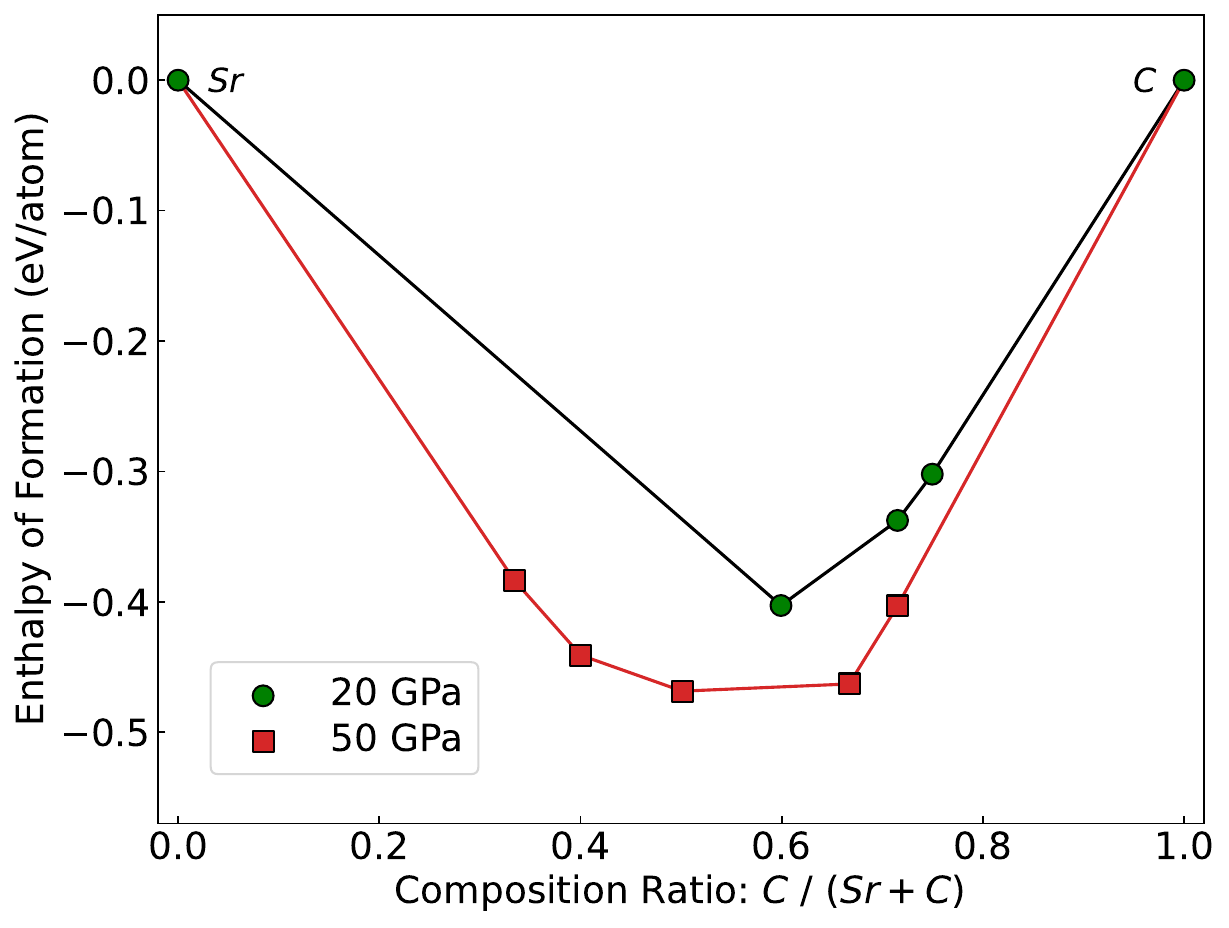}}
	\end{minipage}
    \caption{Convex hull of the Sr-C system at 20~GPa (green disks) and 50~GPa (red squares). The enthalpy of formation $H$ is defined as: $\Delta H(Sr_{x}C_{y})  =  (H(Sr_{x}C_{y}) - xH(Sr) - yH(C)) / (x+y)$.}
    \label{fig:ch}
\end{figure}

\begin{figure}[h!]
	\centering
	\begin{minipage}[h]{1\linewidth}
		\center{\includegraphics[trim={0cm 0cm 0cm 0cm},clip, width=1\linewidth]{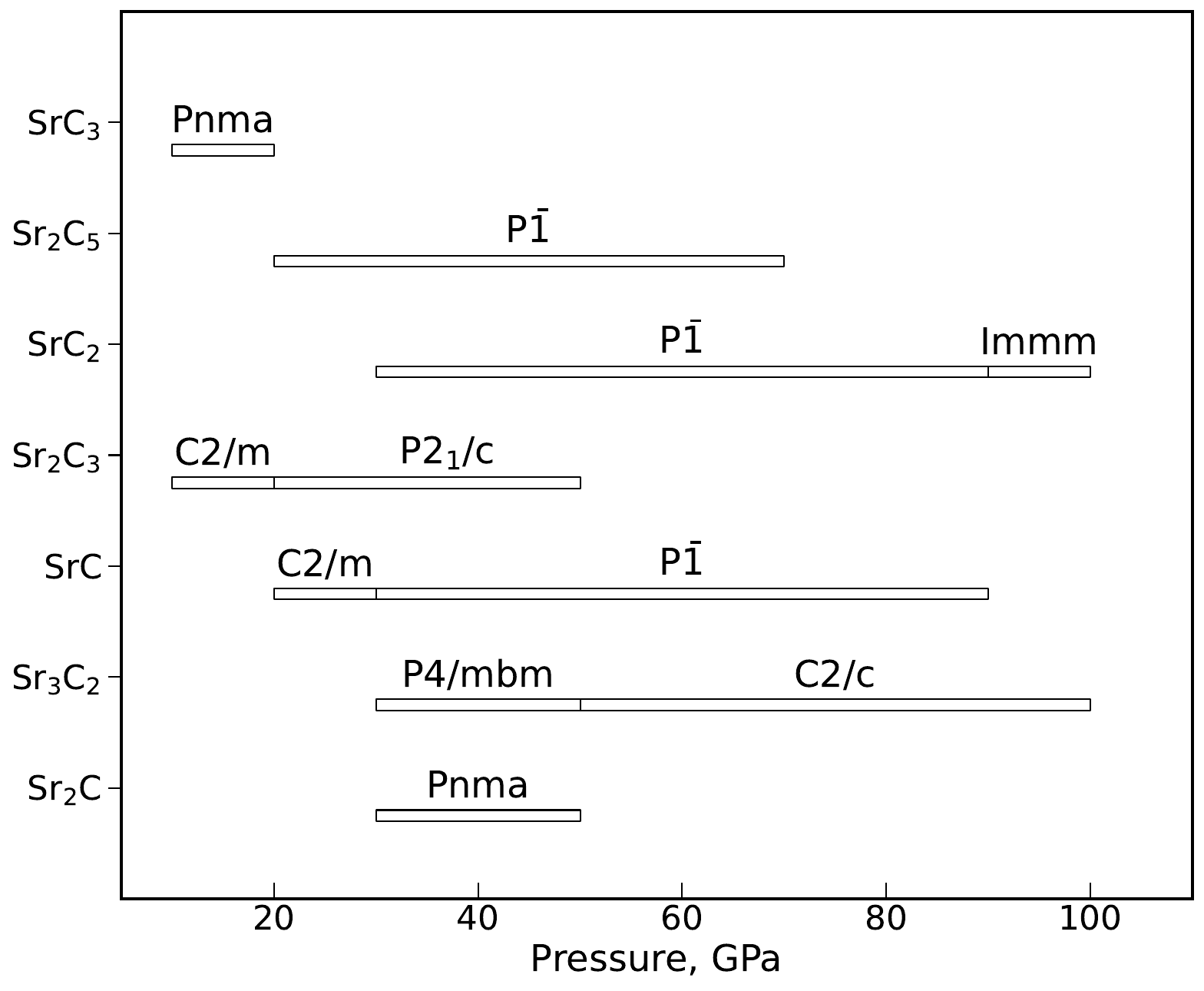}}
	\end{minipage}
    \caption{Pressure-composition phase diagram for the Sr-C system.}
    \label{fig:pd}
\end{figure}

\begin{figure}[h!]
	\centering
	\begin{minipage}[h]{1\linewidth}
		\center{\includegraphics[trim={0cm 5cm 3cm 0cm},clip, width=1\linewidth]{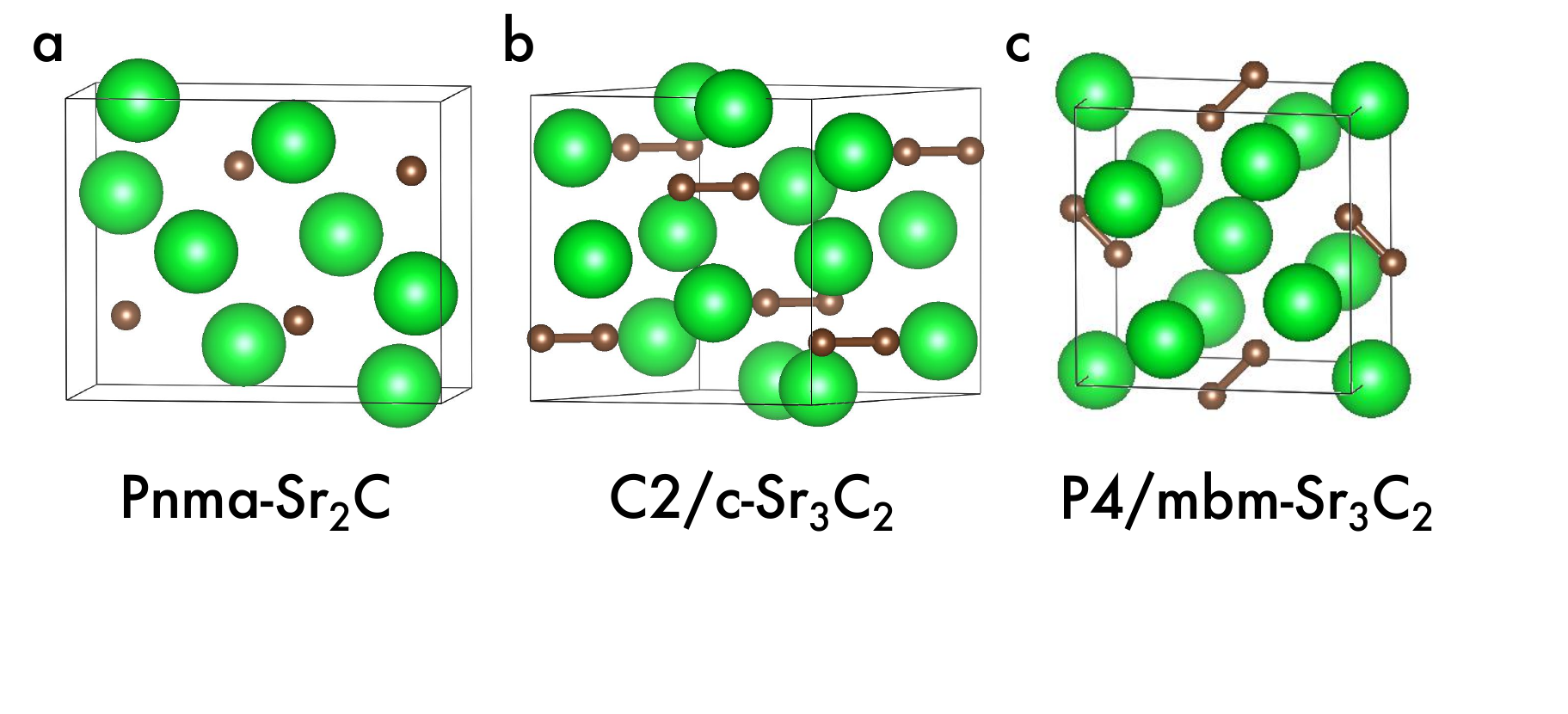}}
	\end{minipage}
 	\begin{minipage}[h]{1\linewidth}
		\center{\includegraphics[trim={0cm 5cm 3cm 0cm},clip, width=1\linewidth]{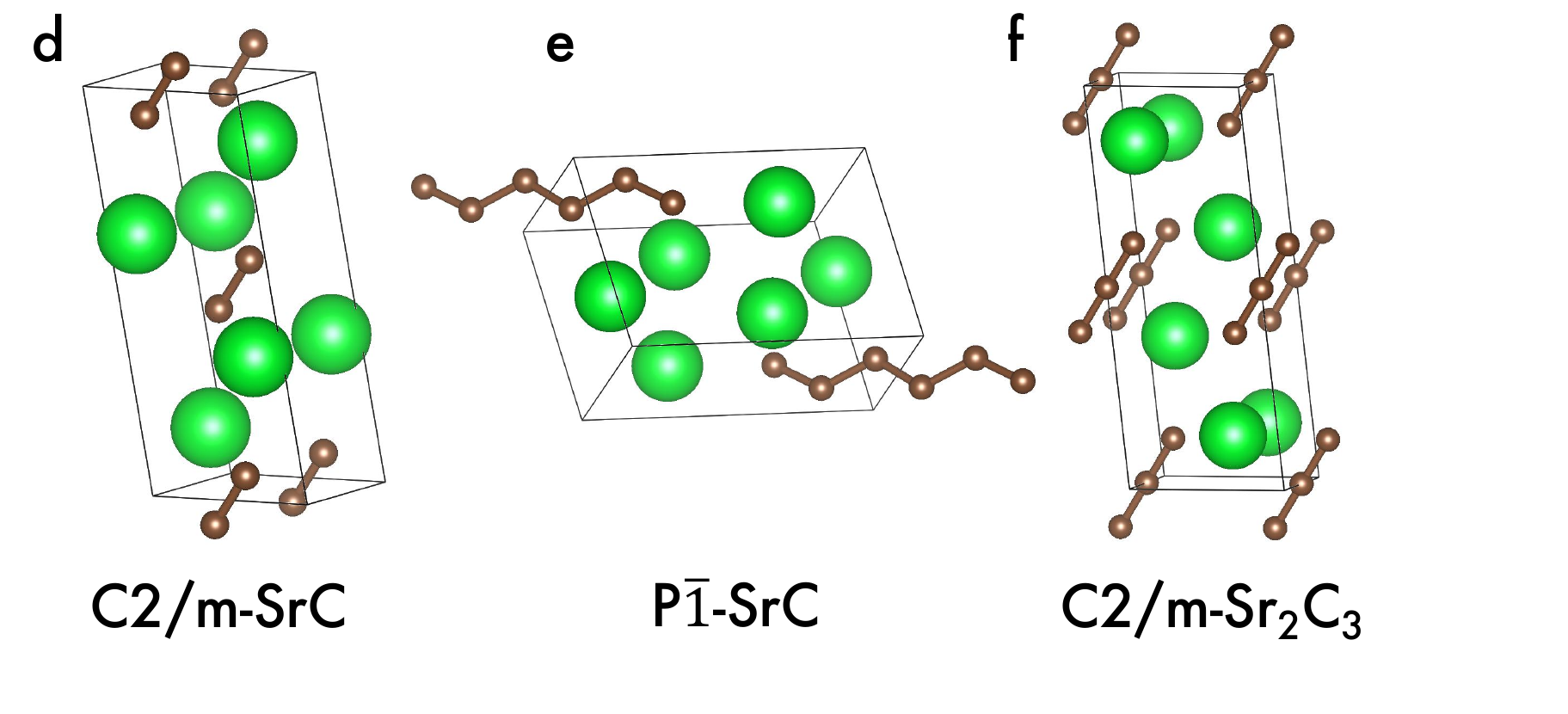}}
	\end{minipage}
 	\begin{minipage}[h]{1\linewidth}
		\center{\includegraphics[trim={0cm 5cm 3cm 0cm},clip, width=1\linewidth]{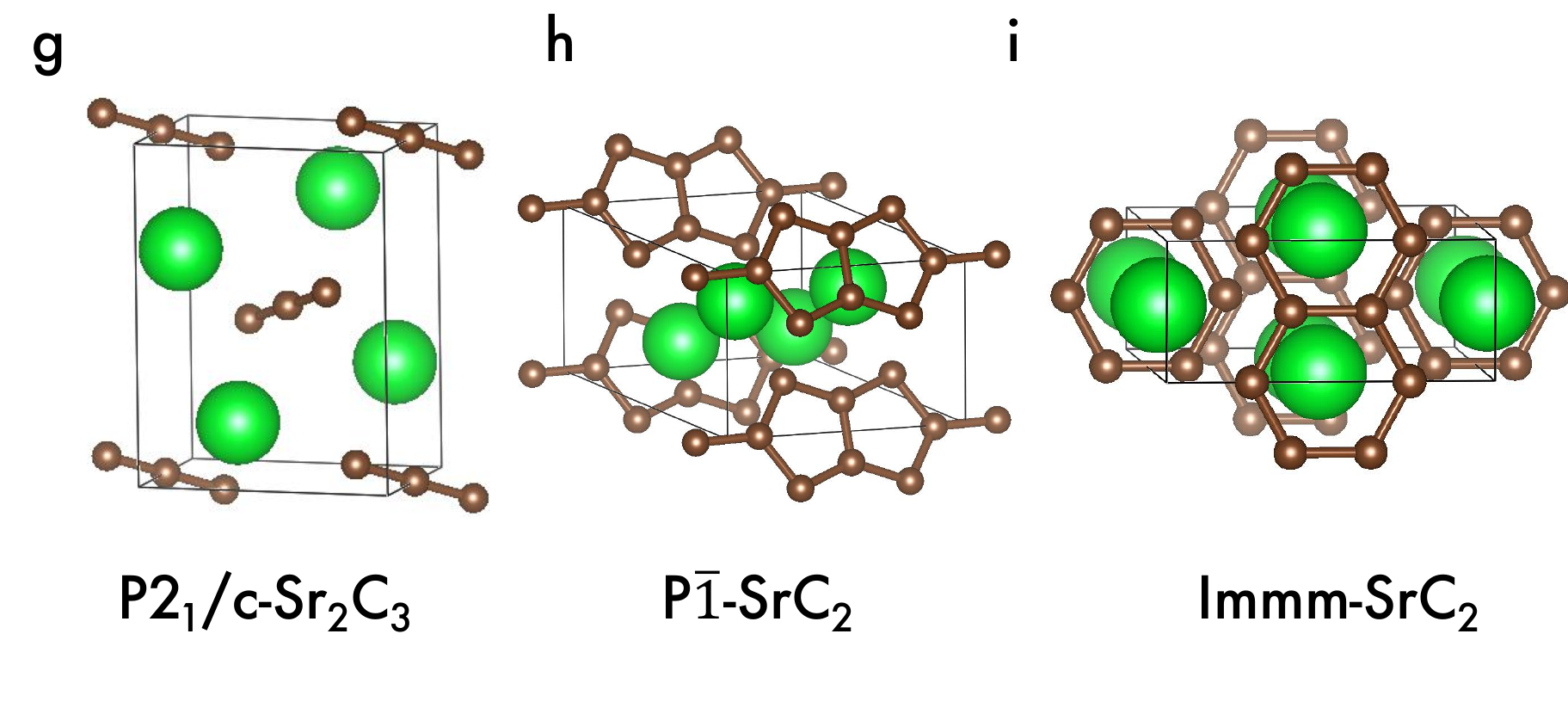}}
	\end{minipage}
  	\begin{minipage}[h]{1\linewidth}
		\center{\includegraphics[trim={0cm 5cm 3cm 0cm},clip, width=1\linewidth]{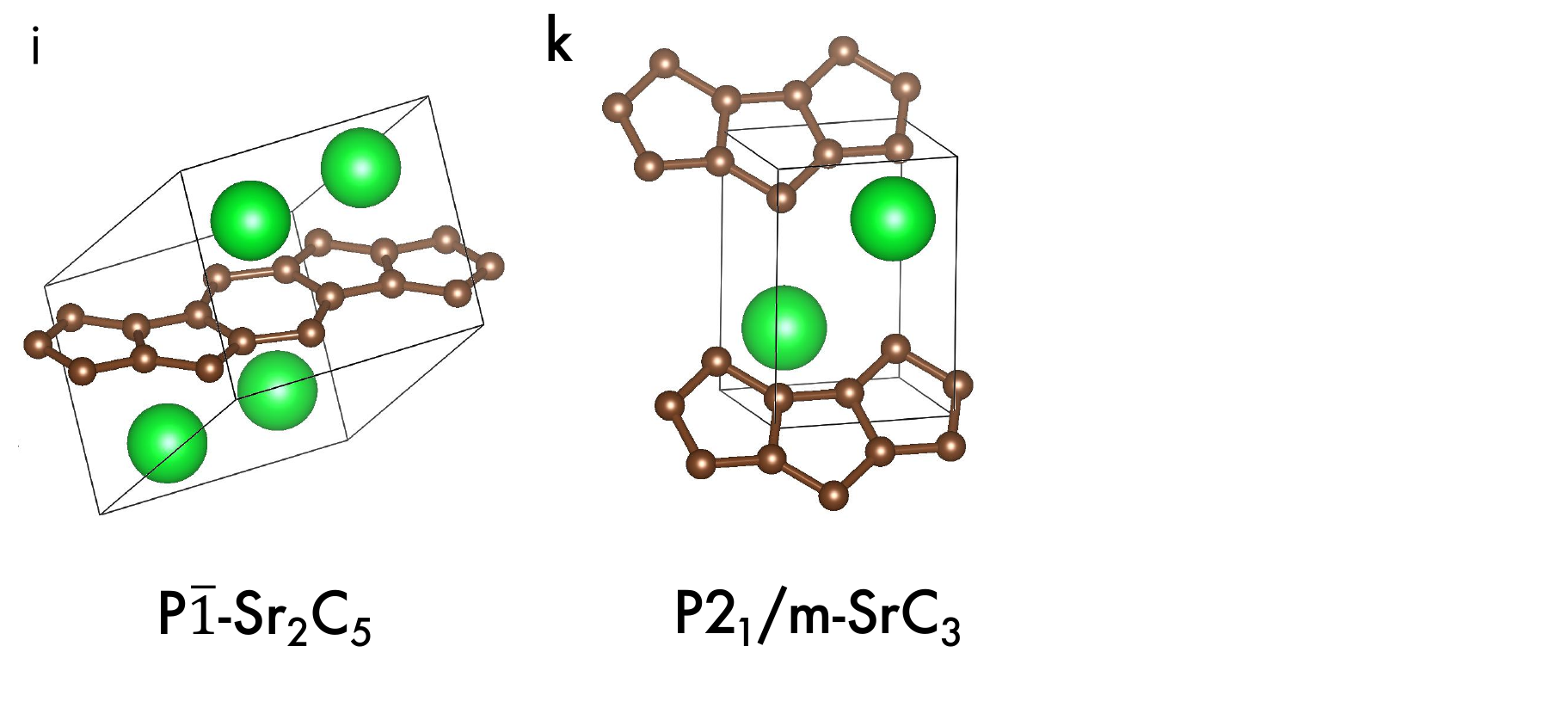}}
	\end{minipage}
	\caption[]{Predicted crystal structures of stable Sr-C compounds. (a) $Pnma$ structure of Sr$_{2}$C. (b, c) $C2/c$ and $P4/mbm$ structures of Sr$_{3}$C$_{2}$. (d, e) $C2/m$ and $P\bar{1}$  structures of SrC. (f, g) $C2/m$ and $P2_{1}/c$ structures of Sr$_{2}$C$_{3}$. (h, i) $P\bar{1}$ and $Immm$ structures of SrC$_{2}$. (j) $P\bar{1}$ structure of Sr$_{2}$C$_{5}$. (k) $P2_{1}/m$ structures of SrC$_{3}$. The green and brown spheres represent strontium and carbon atoms, respectively.}
	\label{fig:src_structures}
\end{figure}

Convex hulls of the Sr-C system at 20~GPa and 50~GPa are presented in Fig.~\ref{fig:ch}. The pressure-composition phase diagram of the Sr-C system is shown in Fig~\ref{fig:pd}, which was obtained by calculating enthalpies of the most stable structures for each composition at a given pressure (steps of 5 GPa were used). According to this diagram, we found eleven hitherto unknown phases in the Sr-C system. These compounds have the following stoichiometries: SrC$_{3}$, Sr$_{2}$C$_{5}$, SrC$_{2}$, Sr$_{2}$C$_{3}$, SrC, Sr$_{3}$C$_{2}$, and Sr$_{2}$C (for some stoichiometries there are a couple of different structures). These structures are shown in Fig.~\ref{fig:src_structures}. For all newly predicted structures, the calculated phonon dispersion relations confirmed their dynamical stability (see SM Fig.S1.).

\begin{figure}[h!]
	\centering
	\begin{minipage}[h]{1\linewidth}
		\center{\includegraphics[trim={0cm 0cm 8cm 0cm},clip, width=1\linewidth]{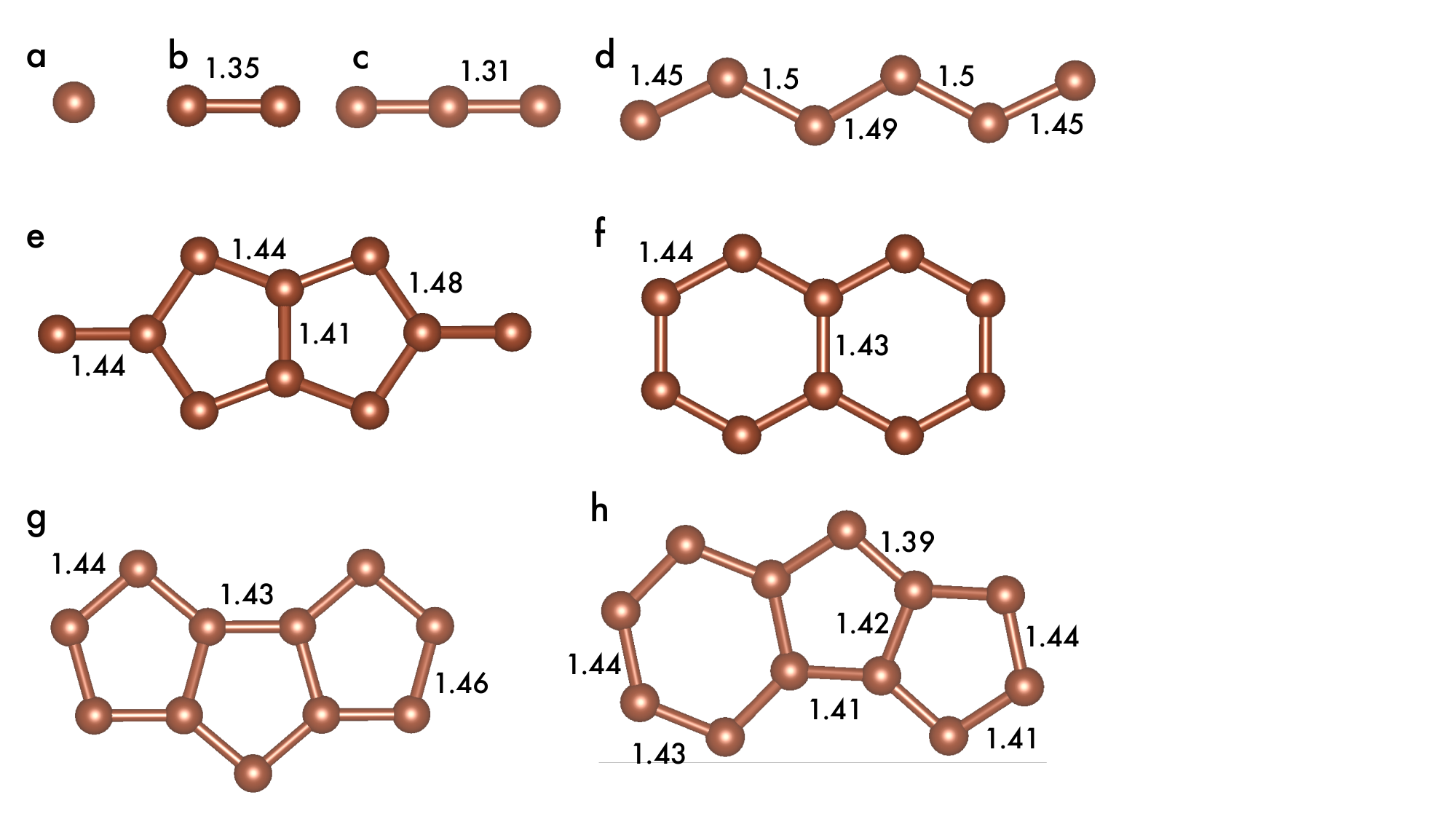}}
	\end{minipage}
	\caption[]{Carbon patterns in the Sr-C system. (a) The isolated carbon anions in the $Pnma$-Sr$_{2}$C. (b) Carbon dimers (dumbbells) observed in the $C2/c$-Sr$_{3}$C$_{2}$ and $C2/m$ structure of SrC. (c) The carbon trimer occurs in the $C2/m$-Sr$_{2}$C$_{3}$. (d) Zigzag C$_6$ groups observed in $P\bar{1}$-SrC at 50 GPa. (e) Carbon stripes in the $P\bar{1}$-SrC$_{2}$ at 50 GPa. (f) Carbon ribbons with hexagonal C-rings in the $Immm$-SrC$_{2}$ at 90 GPa. (g) Carbon ribbons with pentagonal C-rings in the $Pnma$-SrC$_{3}$ at 20 GPa. (h) Carbon ribbons with pentagonal and hexagonal C-rings in the $P\bar{1}$-Sr$_{2}$C$_{5}$ at 50 GPa.}
	\label{fig:src_motifs}
\end{figure}

The identified compounds have a variety of different carbon motifs, as demonstrated in Fig.~\ref{fig:src_motifs}. In the following, we consider the predicted phases in order of increasing carbon content. In order to analyze these structures, we recall that the length of the C - C bond depends on the bond order, and at 1 atm these lengths are 1.20~\r{A} for the triple C–C bond, 1.33~\r{A} for the double and 1.54~\r{A} for the single C–C bond. Combining this knowledge with the results of Bader analysis, we unravel a very diverse chemistry.

According to our calculations Sr$_2$C crystallizes in the orthorhombic $Pnma$ space group. This structure is stable from 30 to 50~GPa, it is metallic, and contains isolated carbon atoms – the simplest carbon motif among all obtained compounds, as shown in Fig.~\ref{fig:src_structures}~(a). Such stoichiometry was previously found in other metal-carbon compounds, for example, Ca$_{2}$C and Mg$_2$C~\cite{li2015investigation,kurakevych2013synthesis}. However, there is no reported theoretical or experimental data on Sr$_{2}$C. This metallic phase (see SM Fig.~S2~(a)) can be represented as a methanide with an idealized charge distribution of (Sr$^{2+}$)$_2$C$^{4-}$ and is in line with the Zintl concept.

\begin{figure}[h!]
	\centering
	\begin{minipage}[h]{1\linewidth}
		\center{\includegraphics[trim={0cm 0cm 0cm 0cm},clip, width=1\linewidth]{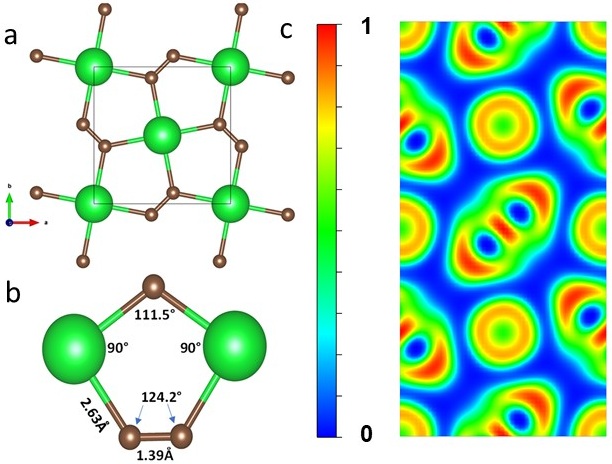}}
	\end{minipage}
    \caption{Crystal structure of $P4/mbm$-Sr$_3$C$_2$  at 30 GPa is depicted along with the computed ELF. Green and brown spheres represent strontium and carbon atoms, respectively. (a) Sr-C plane is a the Cairo pentagonal tiling formed by Sr and C atoms. (b) interatomic distances within the Sr-C plane. (c) 2D ELF of the $P4/mbm$-Sr$_3$C$_2$ structure.}
    \label{fig:src}
\end{figure}

Sr$_3$C$_2$ crystallizes in the tetragonal $P4/mbm$ space group at 30 GPa and transforms into a structure with a monoclinic $C2/c$ space group at 50 GPa. The structure of the newly identified Sr$_3$C$_2$ is of the U$_3$Si$_2$ type, a common arrangement found in silicides~\cite{zachariasen1948crystal}, borides~\cite{riabov1999hydrogenation}, and intermetallic compounds~\cite{chai2011two}. Both structures are metallic and contain carbon dimers as presented in Figs.~\ref{fig:src_structures}~(b,~c). Calcium carbides with this stoichiometry and space groups have been previously predicted~\cite{li2015investigation}. In $P4/mbm$-Sr$_3$C$_2$ phase, the structure contains C$_2$ groups (the C-C distance is 1.39~\r{A} at 30 GPa) carrying an ideal charge of -6, receiving two electrons from three Sr atoms each. This leaves three electrons per C atom to establish C–C double bonds in the metallic framework. A carbon atom with three additional electrons becomes isoelectronic with a fluorine atom, similar to the configuration found in the singly-bonded F$_2$ molecule (where the length of the F-F bond is 1.42~\r{A}). The Sr atoms are located in the same plane as the C$_2$ dumbbells, collectively forming the Cairo pentagonal tiling, which consists of (Sr)$_2$C$_3$ pentagons (see Fig.~\ref{fig:src}~(a)). As shown in Fig.~\ref{fig:src}~(a), Sr atoms are four-fold coordinated by C atoms, with a Sr-C distance of 2.63~\r{A}  at 30 GPa. The ELF indicates a strong covalent bond between the carbon atoms in the dimers and an ionic Sr-C interaction (see Fig.~\ref{fig:src}~(c)). Above 50 GPa the $P4/mbm$-Sr$_3$C$_2$ transforms into $C2/c$-Sr$_3$C$_2$ and the Cairo pentagonal tiling becomes distorted. This distortion is accompanied by an increase in the C-C bond length of the C-dimers to 1.46~\r{A}. Electron density maps in the plane containing the C$_2$ dimers reveal slightly higher ELF values between carbon atoms in the dimers of the $P4/mbm$-Sr$_2$C$_3$ phase compared to the $C2/c$-Sr$_2$C$_3$ phase (see SM Fig.~S6). This suggests a higher bond order in the $P4/mbm$-Sr$_2$C$_3$ dimer.

Metallic SrC exhibits two thermodynamically stable phases below 100 GPa. The low pressure monoclinic $C2/m$ phase on further compression transitions into a low-symmetry triclinic $P\bar{1}$-SrC structure. At 30 GPa the structural formula of $C2/m$ is Sr$_2$(C$_2$), featuring a doubly bonded C$_2$ group with a C–C bond length of 1.34~\r{A}. This C$_2$ group carries an ideal charge of -4 (Bader charge -2.175), perfectly balancing the charge of two strontium atoms, and $C2/m$-SrC is classified as an ethylenide. The $P\bar{1}$-SrC structure is particularly intriguing due to its novel zigzag C-chain with C–C bond lengths ranging from 1.45 to 1.50~\r{A}, implying bond orders between 1 and 2. The bond angles are 125.6$^{\circ}$ and 121.3$^{\circ}$, which are remarkably close to 120$^{\circ}$, expected for sp$^2$-hybridized carbon. The ELF maps clearly indicate strong covalent bonding between the carbon atoms and ionic bonding between Sr and C atoms (see SM Fig.S5). 

At lower pressures of about 10 GPa Sr$_2$C$_3$ becomes stable and crystallizes in the monoclinic $C2/m$ space group. Its structure contains C$_3$ units (allylenide-groups). Beyond 20 GPa, it transforms to the structure with $P2_{1}/c$ space group, which is stable up to 50 GPa. These structures are the only semiconducting compounds found in this study (see SM Fig.~S2). Usually, C$_{3}$ trimers are linear [C=C=C]$^{4-}$ groups with a C-C bond length of 1.34-1.35~\r{A} (e.g., in Mg$_{2}$C$_{3}$), which is close to that in gaseous allene (1.335~\r{A})~\cite{almenningen1959electron}. The structure of $C2/m$-Sr$_2$C$_3$ at 20 GPa consists of Sr$_2$ layers connected by nearly linear, symmetric C$_3$ groups featuring double C-C bonds, with C-C distances measuring 1.31~\r{A}. In this configuration, the ideal charge of the C$_3$ group is -4 (Bader charge -2.625), which precisely balances the total charge of the two Sr atoms in the formula, with one Sr atom having +1.314 Bader charge. The central carbon atom in the C$_3$ group should be neutral and the terminal carbons should have -2 charges as in gaseous allene. The terminal carbon atoms with -2 charges are isoelectronic with oxygen atoms. Therefore, the linear [C=C=C]$^{4-}$ group can structurally behave as [O=C=O], according to Zintl-Klemm rule. However, the central atom of the C$_3$ unit in Sr$_2$C$_3$ exhibits a higher negative Bader charge of -1.225, while the terminal carbon atoms show lower Bader charges of -0.700 (see SM Tab.~1). This discrepancy is attributed to the significant ionic interactions between the Sr atom and the central carbon atom of the C$_3$ unit. This can be seen in the calculated 2D ELF of these structures (see SM Fig.~S3).
	
\begin{figure}[h!]
	\centering
	\begin{minipage}[h]{1\linewidth}
		\center{\includegraphics[trim={0cm 0cm 0cm 0cm},clip, width=1\linewidth]{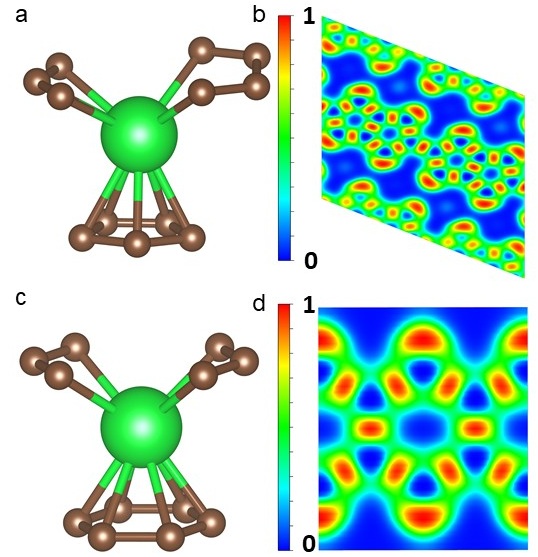}}
	\end{minipage}
    \caption{The crystal structure of SrC$_2$ depicted along with the computed ELF. Green and brown spheres represent strontium and carbon atoms, respectively. (a) Coordination of Sr atoms by C atoms, involving a pentagonal base, (b) 2D ELF containing the pentagonal ribbons, (c) coordination environment of Sr atom in the $Immm$ phase of SrC$_2$ (d) ELF containing the carbon ribbons in the $Immm$-SrC$_2$ phase.}
    \label{fig:SrC2}
\end{figure}

SrC$_2$ becomes stable at 30 GPa and crystallizes in the low-symmetry $P\bar{1}$ space group. This structure contains exotic carbon stripes with fused five-membered C rings -- the structural motif is presented in Fig.~\ref{fig:src_motifs}~(e). In this structure, two layers of Sr atoms are stacked between two layers of these stripes. The C-C distances within the C-rings are in the range from 1.41~\r{A} to 1.48~\r{A} at 30 GPa, indicating bonds, intermediate between single and double. Upon compression, at 90 GPa there is a phase transition $P\bar{1}$-SrC$_2$ $\rightarrow$ $Immm$-SrC$_2$. The $Immm$-SrC$_2$ structure contains extended carbon ribbons consisting of fused hexagonal C-rings as presented in Fig.~\ref{fig:src_motifs}(f). These carbon ribbons are planar, with C–C distances of 1.43~\r{A} and 1.44~\r{A} at 90 GPa. The $Immm$-SrC$_2$ phase shares the same structure as the previously observed high-pressure CaC$_2$ ($Immm$)~\cite{li2015investigation} and the predicted high-pressure YC$_2$ ($Immm$) ~\cite{feng2018carbon}. In the low symmetry $P\bar{1}$-SrC$_2$ phase, the strontium atom is surrounded by 12 carbon atoms that form a heptagon and a pentagon above and below the Sr atom (see Fig.~\ref{fig:SrC2}~(a)). In the $Immm$-SrC$_2$ phase, the strontium atoms have 12 carbon neighbours arranged in a hexagonal prism with unequal bases (see Fig.~\ref{fig:SrC2}~(c)). As was mentioned above, these C atoms are organized into poly-cyclic units of hexagonal rings, forming planar ribbons and as a result, the carbon atoms create exotic, one-dimensional extended poly-anions. The hexagon features two angles of 122.5$^{\circ}$ and four angles of 118.8$^{\circ}$. These angles, being close to 120$^{\circ}$, along with the similar C–C distances, suggest that the carbon atoms are sp$^2$ hybridized. The shared edges of hexagons in $Immm$-SrC$_2$ are slightly shorter (by approximately 0.01~\r{A}) than the non-shared edges. This observation aligns with calculations in previously predicted isostructural high-pressure carbides (such as DyC$_2$, CaC$_2$, and YC$_2$), which show shorter shared edges by about 0.02–0.03~\r{A} ~\cite{akbar2024high}. Bader charge analysis suggests charge transfer from cationic Sr to C atoms (see in SM Tab.~1). ELF maps of carbon ribbons in shown in Figs.~\ref{fig:SrC2}~(b, d) suggest strong covalent bonds between carbon atoms and show localization of electrons on C-C bonds in pentagonal ribbons and hexagons.

\begin{figure}[h!]
	\centering
	\begin{minipage}[h]{1\linewidth}
		\center{\includegraphics[trim={0cm 0cm 0cm 0cm},clip, width=1.0\linewidth]{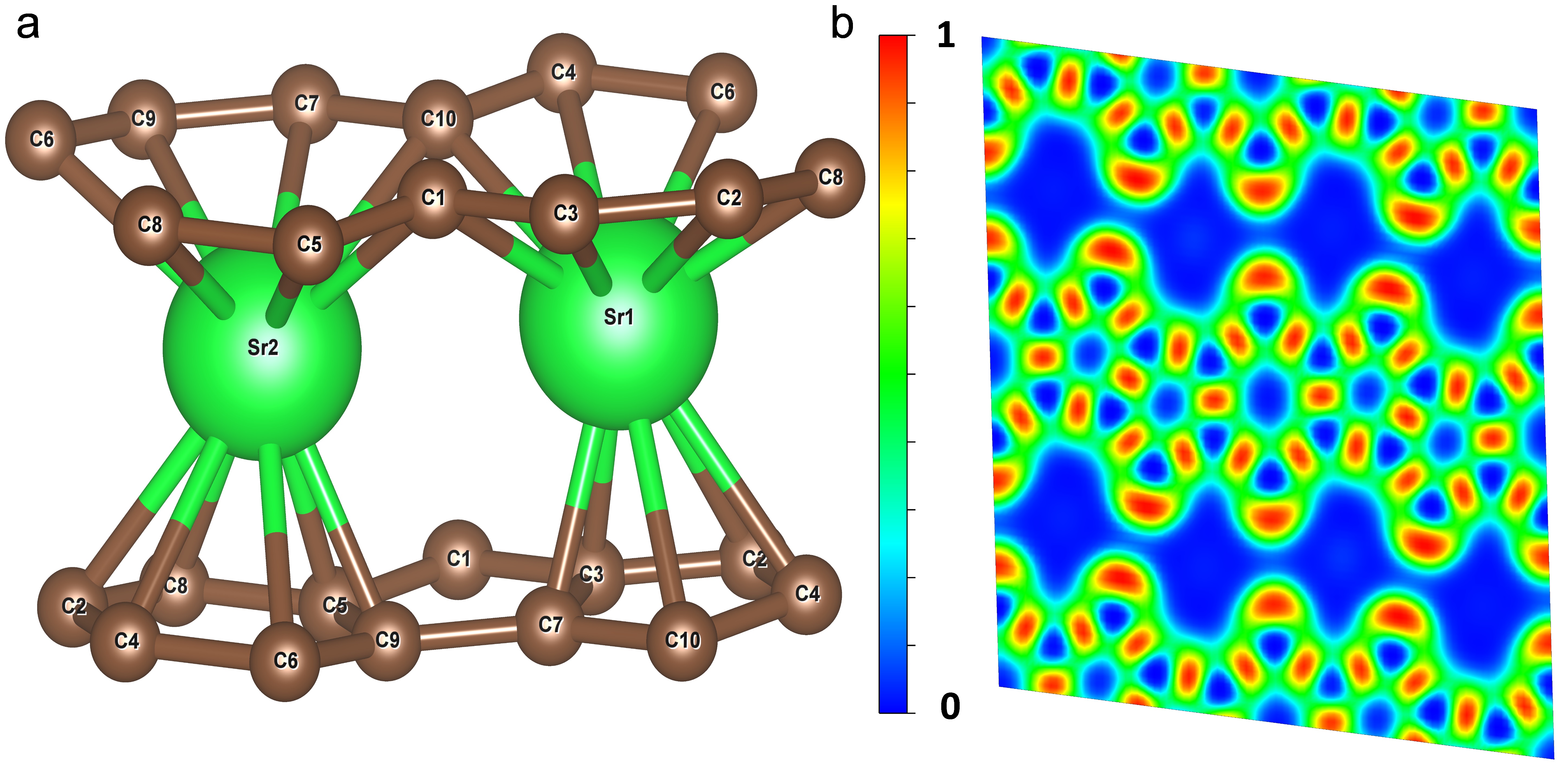}}
	\end{minipage}
    \caption{(a) Coordination environment of the Sr atoms with the fused C-hexagons and C-pentagons in $P\bar{1}$-Sr$_2$C$_5$ (b) 2D ELF of the fused extended C-ribbons, consisting of C-hexagons and C-pentagons}
      \label{fig7:src}
\end{figure}

Sr$_2$C$_5$ crystallizes in the low-symmetry space group $P\bar{1}$. C atoms in this structure feature extended ribbons consisting of C-pentagons and C-hexagons, as shown in Fig.~\ref{fig:src_motifs}(g, h).The C-C bond lengths range from 1.39~\r{A} to 1.44~\r{A}, indicating an intermediate bond order between single and double C-C bonds. In the Sr$_2$C$_5$ structure, one Sr atom has +1.248 Bader charge and is 13-fold coordinated, while another Sr atom has +1.202 Bader charge and is 12-fold coordinated (see Fig.~\ref{fig7:src}(a)). In the fused C-hexagon and C-pentagon, carbon atoms are negatively charged. Therefore, it is evident that there is charge transfer from Sr to C atoms. Sr atom with 13-fold coordination is surrounded by the C-hexagon and a heptagon, while the Sr atom with 12-fold coordination is surrounded by the C-pentagon and a heptagon. In the C-hexagons, the four inner C atoms have three C-C covalent bonds each, while the two outer C atoms have two C-C covalent bonds and one lone pair of electrons each. The C-C bonds in the C-hexagon form two different angles $\alpha$ and $\beta$, where $\alpha$=112$^{\circ}$ and $\beta$= 124$^{\circ}$. In C-pentagons, the inner four C atoms have three C-C covalent bonds, while the outer C atom has two C-C covalent bonds and a lone pair of electrons. The lone pairs of electrons can be seen in the 2-D electron localization function (see Fig.~\ref{fig7:src}~(b)). 

Finally, SrC$_3$ crystallizes in the $Pnma$ space group. In this structure, C atoms form infinite C-ribbons consisting of pentagonal carbon rings (see SM Fig.~S4~(a)). Bader analysis indicates a charge transfer from strontium to carbon atoms, with strontium atoms acquiring a positive Bader charge of +1.343. The carbon atoms, in contrast, have negative charges, with those not involved in ring fusion showing higher electron concentration. This is also evident in the 2D ELF plot (see SM Fig.~S4~(b)).

In light of the appreciable diversity of the observed carbon motifs as demonstrated in Fig.~\ref{fig:src_motifs}, it is instructive to check whether at least some of these
configurations can be explained using conventional chemical wisdom. For that
purpose, we investigated the applicability of the Zintl-Klemm (ZK) concept,
originally conceived to rationalize bonding patterns of polyanions, to the newly identified strontium carbides. To assess the validity of the ZK rules, the average charge of a C-anion will be determined
by assigning the formal $+2$ charge to the Sr atoms (Sr$^{2+}$) and imposing charge neutrality
of the formal Sr$_x$C$_y$ unit. For the example of Sr$_2$C this would lead to a formal
charge of the carbon anion of $-4$. By subsequently equating the structural motifs formed  by that
C-anion to the isoelectronic, neutral species, in the case of C$^{4-}$ that would be a
neon atom, the structure of the polyanion is inferred. Since neon is a noble gas,
and Ne-Ne bonds cannot be formed, we expect no C-C bonds in Sr$_2$C, 
which we indeed observe. We performed this analysis for all herein discussed stoichiometries, which we have summarized in Tab.~\ref{tab:zintl}. We find that for some of the strontium carbide compounds, namely Sr$_2$C, Sr$_3$C$_2$ and SrC, the C-anion structure is consistent with the motifs predicted from the ZK rule. This finding is also in agreement with a previous investigation of high-pressure Ca-C phases~\cite{li2015investigation}.
In the other cases, however, we find that for some
stoichiometries such as SrC$_3$ and Sr$_2$C$_5$ the ZK concept cannot be directly applied. 
Double and triple bonds also lead to exceptions from the ZK rule.
A particularly noteworthy example is the case of 
SrC$_2$, in which the carbon species would be isoelectronic to a nitrogen atom. 
As a consequence, one would expect each carbon atom to form three covalent bonds. 
As illustrated in Fig.~\ref{fig:src_motifs}, we find that the carbon polyanion in $Immm$-SrC$_2$
consists of planar, napthalene-like arrangements. The planarity and the similarity
of all C-C bond lengths in that polyanion suggests that the configuration is aromatic.
This implies that most of the carbon atoms form one single and one double bond,
so that the bonding of these atoms in $Immm$-SrC$_2$ is consistent with the ZK prediction.
The two carbon atoms shared between the two rings, however, must then consequently have two single bonds and one double bond, which is not compatible with the isoelectronic N-atom. The same argument applies to the planar structure of the anion in the $P\overline{1}$ phase of SrC$_2$, so that both SrC$_2$ phases are in contradiction with the ZK rule.

\begin{table}[]
    \centering
    \caption{Applicability of Zintl-Klemm (ZK) rules to the newly identified stoichiometries
    of strontium carbide. In cases where the formal charge was found to be fractional,
    the Zintl-Klemm rules cannot be applied ("N/A") in a straightforward manner.}
    \begin{tabular}{l|l|l|l|l}
    \toprule
        \hline\hline
        \thead{Compound} & \thead{Formal\\C-charge} & \thead{isoelectronic\\to} & \thead{Expected anion\\ structure} & \thead{ZK rule\\fulfilled?}\\
        \midrule
        \hline
         Sr$_2$C     & $-4$    & Ne & mono-atomic & Yes \\
         Sr$_3$C$_2$ & $-3$    & F  & dimer       & Yes \\
         SrC         & $-2$    & O  & dimer       & Yes \\
         Sr$_2$C$_3$ & $-2$    & O  & Linear      & Yes \\
         SrC$_2$     & $-1$    & N  & trigonal non-planar       & No \\
         SrC$_3$     & $-2/3$  & N/A  & N/A       & N/A \\
         Sr$_2$C$_5$ & $-2/5$  & N/A& N/A         & N/A \\
                 \hline\hline
         \bottomrule
    \end{tabular}
    
    \label{tab:zintl}
\end{table}

\section{Conclusion}
In summary, we have produced the first complete pressure-composition phase diagram for Sr-C compounds at pressures up to 100 GPa. As a result we predicted eleven novel thermodynamically stable phases in the Sr-C system. Using electronic structure calculations and by performing Bader charge analysis we unraveled very diverse chemistry in these compounds. Carbon atoms in these compounds form a diverse set of different environments: isolated C atoms in Sr$_{2}$C (methanide), dimers in Sr$_{3}$C$_{2}$ (ethanide) and SrC (ethylenide), linear trimers in Sr$_{2}$C$_{3}$ (allylenide), chains in SrC, stripes with five-membered (C-pentagons) rings in SrC$_{2}$. There are also infinite ribbons consisting of five-membered (C-pentagons) rings in SrC$_{3}$, ribbons consisting of C-pentagons in SrC$_{2}$, and ribbons made of C-pentagons and C-hexagons in Sr$_{2}$C$_{5}$. Bonding patterns in some of the obtained compounds can be described using the Zintl-Klemm rule.

We note that while powerful computational methods, such as evolutionary crystal structure prediction used in this work, are capable of reliably predicting these exotic Sr-C compounds under pressure, experimental validation of the obtained results should be done. 

\section{Acknowledgements}
N.R. and A.S. acknowledge funding from the Russian Science Foundation (Project No. 23-13-00332). A.R.O. and P.G. acknowledge funding from the Russian Science Foundation (Project No. 19-72-30043). E.M. thanks the NOMAD Laboratory at the FHI of the Max Planck Society.

\clearpage
\bibliography{main}

\begin{thebibliography}{68}%
\makeatletter
\providecommand \@ifxundefined [1]{%
 \@ifx{#1\undefined}
}%
\providecommand \@ifnum [1]{%
 \ifnum #1\expandafter \@firstoftwo
 \else \expandafter \@secondoftwo
 \fi
}%
\providecommand \@ifx [1]{%
 \ifx #1\expandafter \@firstoftwo
 \else \expandafter \@secondoftwo
 \fi
}%
\providecommand \natexlab [1]{#1}%
\providecommand \enquote  [1]{``#1''}%
\providecommand \bibnamefont  [1]{#1}%
\providecommand \bibfnamefont [1]{#1}%
\providecommand \citenamefont [1]{#1}%
\providecommand \href@noop [0]{\@secondoftwo}%
\providecommand \href [0]{\begingroup \@sanitize@url \@href}%
\providecommand \@href[1]{\@@startlink{#1}\@@href}%
\providecommand \@@href[1]{\endgroup#1\@@endlink}%
\providecommand \@sanitize@url [0]{\catcode `\\12\catcode `\$12\catcode `\&12\catcode `\#12\catcode `\^12\catcode `\_12\catcode `\%12\relax}%
\providecommand \@@startlink[1]{}%
\providecommand \@@endlink[0]{}%
\providecommand \url  [0]{\begingroup\@sanitize@url \@url }%
\providecommand \@url [1]{\endgroup\@href {#1}{\urlprefix }}%
\providecommand \urlprefix  [0]{URL }%
\providecommand \Eprint [0]{\href }%
\providecommand \doibase [0]{http://dx.doi.org/}%
\providecommand \selectlanguage [0]{\@gobble}%
\providecommand \bibinfo  [0]{\@secondoftwo}%
\providecommand \bibfield  [0]{\@secondoftwo}%
\providecommand \translation [1]{[#1]}%
\providecommand \BibitemOpen [0]{}%
\providecommand \bibitemStop [0]{}%
\providecommand \bibitemNoStop [0]{.\EOS\space}%
\providecommand \EOS [0]{\spacefactor3000\relax}%
\providecommand \BibitemShut  [1]{\csname bibitem#1\endcsname}%
\let\auto@bib@innerbib\@empty
\bibitem [{\citenamefont {McMillan}(2002)}]{mcmillan2002new}%
  \BibitemOpen
  \bibfield  {author} {\bibinfo {author} {\bibfnamefont {P.~F.}\ \bibnamefont {McMillan}},\ }\href@noop {} {\bibfield  {journal} {\bibinfo  {journal} {Nature materials}\ }\textbf {\bibinfo {volume} {1}},\ \bibinfo {pages} {19} (\bibinfo {year} {2002})}\BibitemShut {NoStop}%
\bibitem [{\citenamefont {Grochala}\ \emph {et~al.}(2007)\citenamefont {Grochala}, \citenamefont {Hoffmann}, \citenamefont {Feng},\ and\ \citenamefont {Ashcroft}}]{grochala2007chemical}%
  \BibitemOpen
  \bibfield  {author} {\bibinfo {author} {\bibfnamefont {W.}~\bibnamefont {Grochala}}, \bibinfo {author} {\bibfnamefont {R.}~\bibnamefont {Hoffmann}}, \bibinfo {author} {\bibfnamefont {J.}~\bibnamefont {Feng}}, \ and\ \bibinfo {author} {\bibfnamefont {N.~W.}\ \bibnamefont {Ashcroft}},\ }\href@noop {} {\bibfield  {journal} {\bibinfo  {journal} {Angewandte Chemie International Edition}\ }\textbf {\bibinfo {volume} {46}},\ \bibinfo {pages} {3620} (\bibinfo {year} {2007})}\BibitemShut {NoStop}%
\bibitem [{\citenamefont {Zhang}\ \emph {et~al.}(2013)\citenamefont {Zhang}, \citenamefont {Oganov}, \citenamefont {Goncharov}, \citenamefont {Zhu}, \citenamefont {Boulfelfel}, \citenamefont {Lyakhov}, \citenamefont {Stavrou}, \citenamefont {Somayazulu}, \citenamefont {Prakapenka},\ and\ \citenamefont {Kon{\^o}pkov{\'a}}}]{zhang2013unexpected}%
  \BibitemOpen
  \bibfield  {author} {\bibinfo {author} {\bibfnamefont {W.}~\bibnamefont {Zhang}}, \bibinfo {author} {\bibfnamefont {A.~R.}\ \bibnamefont {Oganov}}, \bibinfo {author} {\bibfnamefont {A.~F.}\ \bibnamefont {Goncharov}}, \bibinfo {author} {\bibfnamefont {Q.}~\bibnamefont {Zhu}}, \bibinfo {author} {\bibfnamefont {S.~E.}\ \bibnamefont {Boulfelfel}}, \bibinfo {author} {\bibfnamefont {A.~O.}\ \bibnamefont {Lyakhov}}, \bibinfo {author} {\bibfnamefont {E.}~\bibnamefont {Stavrou}}, \bibinfo {author} {\bibfnamefont {M.}~\bibnamefont {Somayazulu}}, \bibinfo {author} {\bibfnamefont {V.~B.}\ \bibnamefont {Prakapenka}}, \ and\ \bibinfo {author} {\bibfnamefont {Z.}~\bibnamefont {Kon{\^o}pkov{\'a}}},\ }\href@noop {} {\bibfield  {journal} {\bibinfo  {journal} {Science}\ }\textbf {\bibinfo {volume} {342}},\ \bibinfo {pages} {1502} (\bibinfo {year} {2013})}\BibitemShut {NoStop}%
\bibitem [{\citenamefont {Dong}\ \emph {et~al.}(2017)\citenamefont {Dong}, \citenamefont {Oganov}, \citenamefont {Goncharov}, \citenamefont {Stavrou}, \citenamefont {Lobanov}, \citenamefont {Saleh}, \citenamefont {Qian}, \citenamefont {Zhu}, \citenamefont {Gatti}, \citenamefont {Deringer} \emph {et~al.}}]{dong2017stable}%
  \BibitemOpen
  \bibfield  {author} {\bibinfo {author} {\bibfnamefont {X.}~\bibnamefont {Dong}}, \bibinfo {author} {\bibfnamefont {A.~R.}\ \bibnamefont {Oganov}}, \bibinfo {author} {\bibfnamefont {A.~F.}\ \bibnamefont {Goncharov}}, \bibinfo {author} {\bibfnamefont {E.}~\bibnamefont {Stavrou}}, \bibinfo {author} {\bibfnamefont {S.}~\bibnamefont {Lobanov}}, \bibinfo {author} {\bibfnamefont {G.}~\bibnamefont {Saleh}}, \bibinfo {author} {\bibfnamefont {G.-R.}\ \bibnamefont {Qian}}, \bibinfo {author} {\bibfnamefont {Q.}~\bibnamefont {Zhu}}, \bibinfo {author} {\bibfnamefont {C.}~\bibnamefont {Gatti}}, \bibinfo {author} {\bibfnamefont {V.~L.}\ \bibnamefont {Deringer}},  \emph {et~al.},\ }\href@noop {} {\bibfield  {journal} {\bibinfo  {journal} {Nature Chemistry}\ }\textbf {\bibinfo {volume} {9}},\ \bibinfo {pages} {440} (\bibinfo {year} {2017})}\BibitemShut {NoStop}%
\bibitem [{\citenamefont {Young}\ \emph {et~al.}(2006)\citenamefont {Young}, \citenamefont {Sanloup}, \citenamefont {Gregoryanz}, \citenamefont {Scandolo}, \citenamefont {Hemley},\ and\ \citenamefont {Mao}}]{PhysRevLett.96.155501}%
  \BibitemOpen
  \bibfield  {author} {\bibinfo {author} {\bibfnamefont {A.~F.}\ \bibnamefont {Young}}, \bibinfo {author} {\bibfnamefont {C.}~\bibnamefont {Sanloup}}, \bibinfo {author} {\bibfnamefont {E.}~\bibnamefont {Gregoryanz}}, \bibinfo {author} {\bibfnamefont {S.}~\bibnamefont {Scandolo}}, \bibinfo {author} {\bibfnamefont {R.~J.}\ \bibnamefont {Hemley}}, \ and\ \bibinfo {author} {\bibfnamefont {H.-k.}\ \bibnamefont {Mao}},\ }\href {\doibase 10.1103/PhysRevLett.96.155501} {\bibfield  {journal} {\bibinfo  {journal} {Phys. Rev. Lett.}\ }\textbf {\bibinfo {volume} {96}},\ \bibinfo {pages} {155501} (\bibinfo {year} {2006})}\BibitemShut {NoStop}%
\bibitem [{\citenamefont {Semenok}\ \emph {et~al.}(2020)\citenamefont {Semenok}, \citenamefont {Kruglov}, \citenamefont {Savkin}, \citenamefont {Kvashnin},\ and\ \citenamefont {Oganov}}]{semenok2020distribution}%
  \BibitemOpen
  \bibfield  {author} {\bibinfo {author} {\bibfnamefont {D.~V.}\ \bibnamefont {Semenok}}, \bibinfo {author} {\bibfnamefont {I.~A.}\ \bibnamefont {Kruglov}}, \bibinfo {author} {\bibfnamefont {I.~A.}\ \bibnamefont {Savkin}}, \bibinfo {author} {\bibfnamefont {A.~G.}\ \bibnamefont {Kvashnin}}, \ and\ \bibinfo {author} {\bibfnamefont {A.~R.}\ \bibnamefont {Oganov}},\ }\href@noop {} {\bibfield  {journal} {\bibinfo  {journal} {Current Opinion in Solid State and Materials Science}\ }\textbf {\bibinfo {volume} {24}},\ \bibinfo {pages} {100808} (\bibinfo {year} {2020})}\BibitemShut {NoStop}%
\bibitem [{\citenamefont {Dubrovinsky}\ \emph {et~al.}(2022)\citenamefont {Dubrovinsky}, \citenamefont {Khandarkhaeva}, \citenamefont {Fedotenko}, \citenamefont {Laniel}, \citenamefont {Bykov}, \citenamefont {Giacobbe}, \citenamefont {Lawrence~Bright}, \citenamefont {Sedmak}, \citenamefont {Chariton}, \citenamefont {Prakapenka}, \citenamefont {Ponomareva}, \citenamefont {Smirnova}, \citenamefont {Belov}, \citenamefont {Tasn{\'a}di}, \citenamefont {Shulumba}, \citenamefont {Trybel}, \citenamefont {Abrikosov},\ and\ \citenamefont {Dubrovinskaia}}]{Dubrovinsky2022}%
  \BibitemOpen
  \bibfield  {author} {\bibinfo {author} {\bibfnamefont {L.}~\bibnamefont {Dubrovinsky}}, \bibinfo {author} {\bibfnamefont {S.}~\bibnamefont {Khandarkhaeva}}, \bibinfo {author} {\bibfnamefont {T.}~\bibnamefont {Fedotenko}}, \bibinfo {author} {\bibfnamefont {D.}~\bibnamefont {Laniel}}, \bibinfo {author} {\bibfnamefont {M.}~\bibnamefont {Bykov}}, \bibinfo {author} {\bibfnamefont {C.}~\bibnamefont {Giacobbe}}, \bibinfo {author} {\bibfnamefont {E.}~\bibnamefont {Lawrence~Bright}}, \bibinfo {author} {\bibfnamefont {P.}~\bibnamefont {Sedmak}}, \bibinfo {author} {\bibfnamefont {S.}~\bibnamefont {Chariton}}, \bibinfo {author} {\bibfnamefont {V.}~\bibnamefont {Prakapenka}}, \bibinfo {author} {\bibfnamefont {A.~V.}\ \bibnamefont {Ponomareva}}, \bibinfo {author} {\bibfnamefont {E.~A.}\ \bibnamefont {Smirnova}}, \bibinfo {author} {\bibfnamefont {M.~P.}\ \bibnamefont {Belov}}, \bibinfo {author} {\bibfnamefont {F.}~\bibnamefont {Tasn{\'a}di}}, \bibinfo {author} {\bibfnamefont {N.}~\bibnamefont {Shulumba}}, \bibinfo
  {author} {\bibfnamefont {F.}~\bibnamefont {Trybel}}, \bibinfo {author} {\bibfnamefont {I.~A.}\ \bibnamefont {Abrikosov}}, \ and\ \bibinfo {author} {\bibfnamefont {N.}~\bibnamefont {Dubrovinskaia}},\ }\href {\doibase 10.1038/s41586-022-04550-2} {\bibfield  {journal} {\bibinfo  {journal} {Nature}\ }\textbf {\bibinfo {volume} {605}},\ \bibinfo {pages} {274} (\bibinfo {year} {2022})}\BibitemShut {NoStop}%
\bibitem [{\citenamefont {Oganov}\ \emph {et~al.}(2019)\citenamefont {Oganov}, \citenamefont {Pickard}, \citenamefont {Zhu},\ and\ \citenamefont {Needs}}]{oganov2019structure}%
  \BibitemOpen
  \bibfield  {author} {\bibinfo {author} {\bibfnamefont {A.~R.}\ \bibnamefont {Oganov}}, \bibinfo {author} {\bibfnamefont {C.~J.}\ \bibnamefont {Pickard}}, \bibinfo {author} {\bibfnamefont {Q.}~\bibnamefont {Zhu}}, \ and\ \bibinfo {author} {\bibfnamefont {R.~J.}\ \bibnamefont {Needs}},\ }\href@noop {} {\bibfield  {journal} {\bibinfo  {journal} {Nature Reviews Materials}\ }\textbf {\bibinfo {volume} {4}},\ \bibinfo {pages} {331} (\bibinfo {year} {2019})}\BibitemShut {NoStop}%
\bibitem [{\citenamefont {Miao}\ \emph {et~al.}(2020)\citenamefont {Miao}, \citenamefont {Sun}, \citenamefont {Zurek},\ and\ \citenamefont {Lin}}]{miao2020chemistry}%
  \BibitemOpen
  \bibfield  {author} {\bibinfo {author} {\bibfnamefont {M.}~\bibnamefont {Miao}}, \bibinfo {author} {\bibfnamefont {Y.}~\bibnamefont {Sun}}, \bibinfo {author} {\bibfnamefont {E.}~\bibnamefont {Zurek}}, \ and\ \bibinfo {author} {\bibfnamefont {H.}~\bibnamefont {Lin}},\ }\href@noop {} {\bibfield  {journal} {\bibinfo  {journal} {Nature Reviews Chemistry}\ }\textbf {\bibinfo {volume} {4}},\ \bibinfo {pages} {508} (\bibinfo {year} {2020})}\BibitemShut {NoStop}%
\bibitem [{\citenamefont {Ma}\ \emph {et~al.}(2009)\citenamefont {Ma}, \citenamefont {Eremets}, \citenamefont {Oganov}, \citenamefont {Xie}, \citenamefont {Trojan}, \citenamefont {Medvedev}, \citenamefont {Lyakhov}, \citenamefont {Valle},\ and\ \citenamefont {Prakapenka}}]{ma2009transparent}%
  \BibitemOpen
  \bibfield  {author} {\bibinfo {author} {\bibfnamefont {Y.}~\bibnamefont {Ma}}, \bibinfo {author} {\bibfnamefont {M.}~\bibnamefont {Eremets}}, \bibinfo {author} {\bibfnamefont {A.~R.}\ \bibnamefont {Oganov}}, \bibinfo {author} {\bibfnamefont {Y.}~\bibnamefont {Xie}}, \bibinfo {author} {\bibfnamefont {I.}~\bibnamefont {Trojan}}, \bibinfo {author} {\bibfnamefont {S.}~\bibnamefont {Medvedev}}, \bibinfo {author} {\bibfnamefont {A.~O.}\ \bibnamefont {Lyakhov}}, \bibinfo {author} {\bibfnamefont {M.}~\bibnamefont {Valle}}, \ and\ \bibinfo {author} {\bibfnamefont {V.}~\bibnamefont {Prakapenka}},\ }\href@noop {} {\bibfield  {journal} {\bibinfo  {journal} {Nature}\ }\textbf {\bibinfo {volume} {458}},\ \bibinfo {pages} {182} (\bibinfo {year} {2009})}\BibitemShut {NoStop}%
\bibitem [{\citenamefont {Boeri}\ \emph {et~al.}(2022)\citenamefont {Boeri}, \citenamefont {Hennig}, \citenamefont {Hirschfeld}, \citenamefont {Profeta}, \citenamefont {Sanna}, \citenamefont {Zurek}, \citenamefont {Pickett}, \citenamefont {Amsler}, \citenamefont {Dias}, \citenamefont {Eremets} \emph {et~al.}}]{lilia20222021}%
  \BibitemOpen
  \bibfield  {author} {\bibinfo {author} {\bibfnamefont {L.}~\bibnamefont {Boeri}}, \bibinfo {author} {\bibfnamefont {R.}~\bibnamefont {Hennig}}, \bibinfo {author} {\bibfnamefont {P.}~\bibnamefont {Hirschfeld}}, \bibinfo {author} {\bibfnamefont {G.}~\bibnamefont {Profeta}}, \bibinfo {author} {\bibfnamefont {A.}~\bibnamefont {Sanna}}, \bibinfo {author} {\bibfnamefont {E.}~\bibnamefont {Zurek}}, \bibinfo {author} {\bibfnamefont {W.~E.}\ \bibnamefont {Pickett}}, \bibinfo {author} {\bibfnamefont {M.}~\bibnamefont {Amsler}}, \bibinfo {author} {\bibfnamefont {R.}~\bibnamefont {Dias}}, \bibinfo {author} {\bibfnamefont {M.~I.}\ \bibnamefont {Eremets}},  \emph {et~al.},\ }\href@noop {} {\bibfield  {journal} {\bibinfo  {journal} {Journal of Physics: Condensed Matter}\ }\textbf {\bibinfo {volume} {34}},\ \bibinfo {pages} {183002} (\bibinfo {year} {2022})}\BibitemShut {NoStop}%
\bibitem [{\citenamefont {Zhu}\ \emph {et~al.}(2012)\citenamefont {Zhu}, \citenamefont {Zeng},\ and\ \citenamefont {Oganov}}]{zhu2012systematic}%
  \BibitemOpen
  \bibfield  {author} {\bibinfo {author} {\bibfnamefont {Q.}~\bibnamefont {Zhu}}, \bibinfo {author} {\bibfnamefont {Q.}~\bibnamefont {Zeng}}, \ and\ \bibinfo {author} {\bibfnamefont {A.~R.}\ \bibnamefont {Oganov}},\ }\href@noop {} {\bibfield  {journal} {\bibinfo  {journal} {Physical Review B—Condensed Matter and Materials Physics}\ }\textbf {\bibinfo {volume} {85}},\ \bibinfo {pages} {201407} (\bibinfo {year} {2012})}\BibitemShut {NoStop}%
\bibitem [{\citenamefont {Rice}\ \emph {et~al.}(1983)\citenamefont {Rice}, \citenamefont {Bishop},\ and\ \citenamefont {Campbell}}]{rice1983unusual}%
  \BibitemOpen
  \bibfield  {author} {\bibinfo {author} {\bibfnamefont {M.}~\bibnamefont {Rice}}, \bibinfo {author} {\bibfnamefont {A.}~\bibnamefont {Bishop}}, \ and\ \bibinfo {author} {\bibfnamefont {D.}~\bibnamefont {Campbell}},\ }\href@noop {} {\bibfield  {journal} {\bibinfo  {journal} {Physical review letters}\ }\textbf {\bibinfo {volume} {51}},\ \bibinfo {pages} {2136} (\bibinfo {year} {1983})}\BibitemShut {NoStop}%
\bibitem [{\citenamefont {Hoffmann}\ \emph {et~al.}(1991)\citenamefont {Hoffmann}, \citenamefont {Janiak},\ and\ \citenamefont {Kollmar}}]{hoffmann1991chemical}%
  \BibitemOpen
  \bibfield  {author} {\bibinfo {author} {\bibfnamefont {R.}~\bibnamefont {Hoffmann}}, \bibinfo {author} {\bibfnamefont {C.}~\bibnamefont {Janiak}}, \ and\ \bibinfo {author} {\bibfnamefont {C.}~\bibnamefont {Kollmar}},\ }\href@noop {} {\bibfield  {journal} {\bibinfo  {journal} {Macromolecules}\ }\textbf {\bibinfo {volume} {24}},\ \bibinfo {pages} {3725} (\bibinfo {year} {1991})}\BibitemShut {NoStop}%
\bibitem [{\citenamefont {Wang}\ \emph {et~al.}(2015)\citenamefont {Wang}, \citenamefont {Zhou}, \citenamefont {Zhang}, \citenamefont {Zhu}, \citenamefont {Dong}, \citenamefont {Zhao},\ and\ \citenamefont {Oganov}}]{wang2015phagraphene}%
  \BibitemOpen
  \bibfield  {author} {\bibinfo {author} {\bibfnamefont {Z.}~\bibnamefont {Wang}}, \bibinfo {author} {\bibfnamefont {X.-F.}\ \bibnamefont {Zhou}}, \bibinfo {author} {\bibfnamefont {X.}~\bibnamefont {Zhang}}, \bibinfo {author} {\bibfnamefont {Q.}~\bibnamefont {Zhu}}, \bibinfo {author} {\bibfnamefont {H.}~\bibnamefont {Dong}}, \bibinfo {author} {\bibfnamefont {M.}~\bibnamefont {Zhao}}, \ and\ \bibinfo {author} {\bibfnamefont {A.~R.}\ \bibnamefont {Oganov}},\ }\href@noop {} {\bibfield  {journal} {\bibinfo  {journal} {Nano letters}\ }\textbf {\bibinfo {volume} {15}},\ \bibinfo {pages} {6182} (\bibinfo {year} {2015})}\BibitemShut {NoStop}%
\bibitem [{\citenamefont {Li}\ \emph {et~al.}(2009)\citenamefont {Li}, \citenamefont {Ma}, \citenamefont {Oganov}, \citenamefont {Wang}, \citenamefont {Wang}, \citenamefont {Xu}, \citenamefont {Cui}, \citenamefont {Mao},\ and\ \citenamefont {Zou}}]{li2009superhard}%
  \BibitemOpen
  \bibfield  {author} {\bibinfo {author} {\bibfnamefont {Q.}~\bibnamefont {Li}}, \bibinfo {author} {\bibfnamefont {Y.}~\bibnamefont {Ma}}, \bibinfo {author} {\bibfnamefont {A.~R.}\ \bibnamefont {Oganov}}, \bibinfo {author} {\bibfnamefont {H.}~\bibnamefont {Wang}}, \bibinfo {author} {\bibfnamefont {H.}~\bibnamefont {Wang}}, \bibinfo {author} {\bibfnamefont {Y.}~\bibnamefont {Xu}}, \bibinfo {author} {\bibfnamefont {.~f.~T.}\ \bibnamefont {Cui}}, \bibinfo {author} {\bibfnamefont {H.-K.}\ \bibnamefont {Mao}}, \ and\ \bibinfo {author} {\bibfnamefont {G.}~\bibnamefont {Zou}},\ }\href@noop {} {\bibfield  {journal} {\bibinfo  {journal} {Physical review letters}\ }\textbf {\bibinfo {volume} {102}},\ \bibinfo {pages} {175506} (\bibinfo {year} {2009})}\BibitemShut {NoStop}%
\bibitem [{\citenamefont {Boulfelfel}\ \emph {et~al.}(2012)\citenamefont {Boulfelfel}, \citenamefont {Oganov},\ and\ \citenamefont {Leoni}}]{boulfelfel2012understanding}%
  \BibitemOpen
  \bibfield  {author} {\bibinfo {author} {\bibfnamefont {S.~E.}\ \bibnamefont {Boulfelfel}}, \bibinfo {author} {\bibfnamefont {A.~R.}\ \bibnamefont {Oganov}}, \ and\ \bibinfo {author} {\bibfnamefont {S.}~\bibnamefont {Leoni}},\ }\href@noop {} {\bibfield  {journal} {\bibinfo  {journal} {Scientific Reports}\ }\textbf {\bibinfo {volume} {2}},\ \bibinfo {pages} {471} (\bibinfo {year} {2012})}\BibitemShut {NoStop}%
\bibitem [{\citenamefont {Wang}\ \emph {et~al.}(2012)\citenamefont {Wang}, \citenamefont {Panzik}, \citenamefont {Kiefer},\ and\ \citenamefont {Lee}}]{wang2012crystal}%
  \BibitemOpen
  \bibfield  {author} {\bibinfo {author} {\bibfnamefont {Y.}~\bibnamefont {Wang}}, \bibinfo {author} {\bibfnamefont {J.~E.}\ \bibnamefont {Panzik}}, \bibinfo {author} {\bibfnamefont {B.}~\bibnamefont {Kiefer}}, \ and\ \bibinfo {author} {\bibfnamefont {K.~K.}\ \bibnamefont {Lee}},\ }\href@noop {} {\bibfield  {journal} {\bibinfo  {journal} {Scientific reports}\ }\textbf {\bibinfo {volume} {2}},\ \bibinfo {pages} {520} (\bibinfo {year} {2012})}\BibitemShut {NoStop}%
\bibitem [{\citenamefont {Hoffmann}\ and\ \citenamefont {Meyer}(1992)}]{hoffmann1992electronic}%
  \BibitemOpen
  \bibfield  {author} {\bibinfo {author} {\bibfnamefont {R.}~\bibnamefont {Hoffmann}}\ and\ \bibinfo {author} {\bibfnamefont {H.-J.}\ \bibnamefont {Meyer}},\ }\href@noop {} {\bibfield  {journal} {\bibinfo  {journal} {Zeitschrift f{\"u}r anorganische und allgemeine Chemie}\ }\textbf {\bibinfo {volume} {607}},\ \bibinfo {pages} {57} (\bibinfo {year} {1992})}\BibitemShut {NoStop}%
\bibitem [{\citenamefont {Su}\ \emph {et~al.}(2016)\citenamefont {Su}, \citenamefont {Zhang}, \citenamefont {Liu}, \citenamefont {Wang}, \citenamefont {Wang},\ and\ \citenamefont {Ma}}]{su2016catenation}%
  \BibitemOpen
  \bibfield  {author} {\bibinfo {author} {\bibfnamefont {C.}~\bibnamefont {Su}}, \bibinfo {author} {\bibfnamefont {J.}~\bibnamefont {Zhang}}, \bibinfo {author} {\bibfnamefont {G.}~\bibnamefont {Liu}}, \bibinfo {author} {\bibfnamefont {X.}~\bibnamefont {Wang}}, \bibinfo {author} {\bibfnamefont {H.}~\bibnamefont {Wang}}, \ and\ \bibinfo {author} {\bibfnamefont {Y.}~\bibnamefont {Ma}},\ }\href@noop {} {\bibfield  {journal} {\bibinfo  {journal} {Physical Chemistry Chemical Physics}\ }\textbf {\bibinfo {volume} {18}},\ \bibinfo {pages} {14286} (\bibinfo {year} {2016})}\BibitemShut {NoStop}%
\bibitem [{\citenamefont {Krupka}\ \emph {et~al.}(1969)\citenamefont {Krupka}, \citenamefont {Giorgi}, \citenamefont {Krikorian},\ and\ \citenamefont {Szklarz}}]{krupka1969high}%
  \BibitemOpen
  \bibfield  {author} {\bibinfo {author} {\bibfnamefont {M.}~\bibnamefont {Krupka}}, \bibinfo {author} {\bibfnamefont {A.}~\bibnamefont {Giorgi}}, \bibinfo {author} {\bibfnamefont {N.}~\bibnamefont {Krikorian}}, \ and\ \bibinfo {author} {\bibfnamefont {E.}~\bibnamefont {Szklarz}},\ }\href@noop {} {\bibfield  {journal} {\bibinfo  {journal} {Journal of the Less Common Metals}\ }\textbf {\bibinfo {volume} {17}},\ \bibinfo {pages} {91} (\bibinfo {year} {1969})}\BibitemShut {NoStop}%
\bibitem [{\citenamefont {Parasuk}\ and\ \citenamefont {Alml{\"o}f}(1989)}]{parasuk1989electronic}%
  \BibitemOpen
  \bibfield  {author} {\bibinfo {author} {\bibfnamefont {V.}~\bibnamefont {Parasuk}}\ and\ \bibinfo {author} {\bibfnamefont {J.}~\bibnamefont {Alml{\"o}f}},\ }\href@noop {} {\bibfield  {journal} {\bibinfo  {journal} {The Journal of chemical physics}\ }\textbf {\bibinfo {volume} {91}},\ \bibinfo {pages} {1137} (\bibinfo {year} {1989})}\BibitemShut {NoStop}%
\bibitem [{\citenamefont {Toth}(2014)}]{toth2014transition}%
  \BibitemOpen
  \bibfield  {author} {\bibinfo {author} {\bibfnamefont {L.}~\bibnamefont {Toth}},\ }\href@noop {} {\emph {\bibinfo {title} {Transition metal carbides and nitrides}}}\ (\bibinfo  {publisher} {Elsevier},\ \bibinfo {year} {2014})\BibitemShut {NoStop}%
\bibitem [{\citenamefont {Spedding}\ \emph {et~al.}(1958)\citenamefont {Spedding}, \citenamefont {Gschneidner~Jr},\ and\ \citenamefont {Daane}}]{spedding1958crystal}%
  \BibitemOpen
  \bibfield  {author} {\bibinfo {author} {\bibfnamefont {F.}~\bibnamefont {Spedding}}, \bibinfo {author} {\bibfnamefont {K.}~\bibnamefont {Gschneidner~Jr}}, \ and\ \bibinfo {author} {\bibfnamefont {A.}~\bibnamefont {Daane}},\ }\href@noop {} {\bibfield  {journal} {\bibinfo  {journal} {Journal of the American Chemical Society}\ }\textbf {\bibinfo {volume} {80}},\ \bibinfo {pages} {4499} (\bibinfo {year} {1958})}\BibitemShut {NoStop}%
\bibitem [{\citenamefont {Sakai}\ \emph {et~al.}(1981)\citenamefont {Sakai}, \citenamefont {Adachi}, \citenamefont {Yoshida},\ and\ \citenamefont {Shiokawa}}]{sakai1981magnetic}%
  \BibitemOpen
  \bibfield  {author} {\bibinfo {author} {\bibfnamefont {T.}~\bibnamefont {Sakai}}, \bibinfo {author} {\bibfnamefont {G.-y.}\ \bibnamefont {Adachi}}, \bibinfo {author} {\bibfnamefont {T.}~\bibnamefont {Yoshida}}, \ and\ \bibinfo {author} {\bibfnamefont {J.}~\bibnamefont {Shiokawa}},\ }\href@noop {} {\bibfield  {journal} {\bibinfo  {journal} {The Journal of Chemical Physics}\ }\textbf {\bibinfo {volume} {75}},\ \bibinfo {pages} {3027} (\bibinfo {year} {1981})}\BibitemShut {NoStop}%
\bibitem [{\citenamefont {Rundle}\ \emph {et~al.}(1948)\citenamefont {Rundle}, \citenamefont {Baenziger}, \citenamefont {Wilson},\ and\ \citenamefont {McDonald}}]{rundle1948structures}%
  \BibitemOpen
  \bibfield  {author} {\bibinfo {author} {\bibfnamefont {R.}~\bibnamefont {Rundle}}, \bibinfo {author} {\bibfnamefont {N.}~\bibnamefont {Baenziger}}, \bibinfo {author} {\bibfnamefont {A.}~\bibnamefont {Wilson}}, \ and\ \bibinfo {author} {\bibfnamefont {R.}~\bibnamefont {McDonald}},\ }\href@noop {} {\bibfield  {journal} {\bibinfo  {journal} {Journal of the American Chemical Society}\ }\textbf {\bibinfo {volume} {70}},\ \bibinfo {pages} {99} (\bibinfo {year} {1948})}\BibitemShut {NoStop}%
\bibitem [{\citenamefont {Akbar}\ \emph {et~al.}(2024{\natexlab{a}})\citenamefont {Akbar}, \citenamefont {Aslandukova}, \citenamefont {Yin}, \citenamefont {Aslandukov}, \citenamefont {Laniel}, \citenamefont {Bykova}, \citenamefont {Bykov}, \citenamefont {Bright}, \citenamefont {Wright}, \citenamefont {Comboni}, \citenamefont {Hanfland}, \citenamefont {Dubrovinskaia},\ and\ \citenamefont {Dubrovinsky}}]{AKBAR2024119374}%
  \BibitemOpen
  \bibfield  {author} {\bibinfo {author} {\bibfnamefont {F.~I.}\ \bibnamefont {Akbar}}, \bibinfo {author} {\bibfnamefont {A.}~\bibnamefont {Aslandukova}}, \bibinfo {author} {\bibfnamefont {Y.}~\bibnamefont {Yin}}, \bibinfo {author} {\bibfnamefont {A.}~\bibnamefont {Aslandukov}}, \bibinfo {author} {\bibfnamefont {D.}~\bibnamefont {Laniel}}, \bibinfo {author} {\bibfnamefont {E.}~\bibnamefont {Bykova}}, \bibinfo {author} {\bibfnamefont {M.}~\bibnamefont {Bykov}}, \bibinfo {author} {\bibfnamefont {E.~L.}\ \bibnamefont {Bright}}, \bibinfo {author} {\bibfnamefont {J.}~\bibnamefont {Wright}}, \bibinfo {author} {\bibfnamefont {D.}~\bibnamefont {Comboni}}, \bibinfo {author} {\bibfnamefont {M.}~\bibnamefont {Hanfland}}, \bibinfo {author} {\bibfnamefont {N.}~\bibnamefont {Dubrovinskaia}}, \ and\ \bibinfo {author} {\bibfnamefont {L.}~\bibnamefont {Dubrovinsky}},\ }\href {\doibase https://doi.org/10.1016/j.carbon.2024.119374} {\bibfield  {journal} {\bibinfo  {journal} {Carbon}\ }\textbf {\bibinfo {volume} {228}},\
  \bibinfo {pages} {119374} (\bibinfo {year} {2024}{\natexlab{a}})}\BibitemShut {NoStop}%
\bibitem [{\citenamefont {Feng}\ \emph {et~al.}(2018)\citenamefont {Feng}, \citenamefont {Lu}, \citenamefont {Pickard}, \citenamefont {Liu}, \citenamefont {Redfern},\ and\ \citenamefont {Ma}}]{feng2018carbon}%
  \BibitemOpen
  \bibfield  {author} {\bibinfo {author} {\bibfnamefont {X.}~\bibnamefont {Feng}}, \bibinfo {author} {\bibfnamefont {S.}~\bibnamefont {Lu}}, \bibinfo {author} {\bibfnamefont {C.~J.}\ \bibnamefont {Pickard}}, \bibinfo {author} {\bibfnamefont {H.}~\bibnamefont {Liu}}, \bibinfo {author} {\bibfnamefont {S.~A.}\ \bibnamefont {Redfern}}, \ and\ \bibinfo {author} {\bibfnamefont {Y.}~\bibnamefont {Ma}},\ }\href@noop {} {\bibfield  {journal} {\bibinfo  {journal} {Communications Chemistry}\ }\textbf {\bibinfo {volume} {1}},\ \bibinfo {pages} {85} (\bibinfo {year} {2018})}\BibitemShut {NoStop}%
\bibitem [{\citenamefont {West}\ \emph {et~al.}(1965)\citenamefont {West}, \citenamefont {Carney},\ and\ \citenamefont {Mineo}}]{west1965tetralithium}%
  \BibitemOpen
  \bibfield  {author} {\bibinfo {author} {\bibfnamefont {R.}~\bibnamefont {West}}, \bibinfo {author} {\bibfnamefont {P.~A.}\ \bibnamefont {Carney}}, \ and\ \bibinfo {author} {\bibfnamefont {I.}~\bibnamefont {Mineo}},\ }\href@noop {} {\bibfield  {journal} {\bibinfo  {journal} {Journal of the American Chemical Society}\ }\textbf {\bibinfo {volume} {87}},\ \bibinfo {pages} {3788} (\bibinfo {year} {1965})}\BibitemShut {NoStop}%
\bibitem [{\citenamefont {Mattausch}\ \emph {et~al.}(1994)\citenamefont {Mattausch}, \citenamefont {Gulden}, \citenamefont {Kremer}, \citenamefont {Horakh},\ and\ \citenamefont {Simon}}]{mattausch1994ho4c7}%
  \BibitemOpen
  \bibfield  {author} {\bibinfo {author} {\bibfnamefont {H.}~\bibnamefont {Mattausch}}, \bibinfo {author} {\bibfnamefont {T.}~\bibnamefont {Gulden}}, \bibinfo {author} {\bibfnamefont {R.~K.}\ \bibnamefont {Kremer}}, \bibinfo {author} {\bibfnamefont {J.}~\bibnamefont {Horakh}}, \ and\ \bibinfo {author} {\bibfnamefont {A.}~\bibnamefont {Simon}},\ }\href@noop {} {\bibfield  {journal} {\bibinfo  {journal} {Zeitschrift f{\"u}r Naturforschung B}\ }\textbf {\bibinfo {volume} {49}},\ \bibinfo {pages} {1439} (\bibinfo {year} {1994})}\BibitemShut {NoStop}%
\bibitem [{\citenamefont {Fjellvaag}\ and\ \citenamefont {Karen}(1992)}]{fjellvaag1992crystal}%
  \BibitemOpen
  \bibfield  {author} {\bibinfo {author} {\bibfnamefont {H.}~\bibnamefont {Fjellvaag}}\ and\ \bibinfo {author} {\bibfnamefont {P.}~\bibnamefont {Karen}},\ }\href@noop {} {\bibfield  {journal} {\bibinfo  {journal} {Inorganic Chemistry}\ }\textbf {\bibinfo {volume} {31}},\ \bibinfo {pages} {3260} (\bibinfo {year} {1992})}\BibitemShut {NoStop}%
\bibitem [{\citenamefont {Poettgen}\ and\ \citenamefont {Jeitschko}(1991)}]{poettgen1991scandium}%
  \BibitemOpen
  \bibfield  {author} {\bibinfo {author} {\bibfnamefont {R.}~\bibnamefont {Poettgen}}\ and\ \bibinfo {author} {\bibfnamefont {W.}~\bibnamefont {Jeitschko}},\ }\href@noop {} {\bibfield  {journal} {\bibinfo  {journal} {Inorganic Chemistry}\ }\textbf {\bibinfo {volume} {30}},\ \bibinfo {pages} {427} (\bibinfo {year} {1991})}\BibitemShut {NoStop}%
\bibitem [{\citenamefont {Strobel}\ \emph {et~al.}(2014)\citenamefont {Strobel}, \citenamefont {Kurakevych}, \citenamefont {Kim}, \citenamefont {Le~Godec}, \citenamefont {Crichton}, \citenamefont {Guignard}, \citenamefont {Guignot}, \citenamefont {Cody},\ and\ \citenamefont {Oganov}}]{strobel2014synthesis}%
  \BibitemOpen
  \bibfield  {author} {\bibinfo {author} {\bibfnamefont {T.~A.}\ \bibnamefont {Strobel}}, \bibinfo {author} {\bibfnamefont {O.~O.}\ \bibnamefont {Kurakevych}}, \bibinfo {author} {\bibfnamefont {D.~Y.}\ \bibnamefont {Kim}}, \bibinfo {author} {\bibfnamefont {Y.}~\bibnamefont {Le~Godec}}, \bibinfo {author} {\bibfnamefont {W.}~\bibnamefont {Crichton}}, \bibinfo {author} {\bibfnamefont {J.}~\bibnamefont {Guignard}}, \bibinfo {author} {\bibfnamefont {N.}~\bibnamefont {Guignot}}, \bibinfo {author} {\bibfnamefont {G.~D.}\ \bibnamefont {Cody}}, \ and\ \bibinfo {author} {\bibfnamefont {A.~R.}\ \bibnamefont {Oganov}},\ }\href@noop {} {\bibfield  {journal} {\bibinfo  {journal} {Inorganic Chemistry}\ }\textbf {\bibinfo {volume} {53}},\ \bibinfo {pages} {7020} (\bibinfo {year} {2014})}\BibitemShut {NoStop}%
\bibitem [{\citenamefont {Li}\ \emph {et~al.}(2015)\citenamefont {Li}, \citenamefont {Wang}, \citenamefont {Oganov}, \citenamefont {Gou}, \citenamefont {Smith},\ and\ \citenamefont {Strobel}}]{li2015investigation}%
  \BibitemOpen
  \bibfield  {author} {\bibinfo {author} {\bibfnamefont {Y.-L.}\ \bibnamefont {Li}}, \bibinfo {author} {\bibfnamefont {S.-N.}\ \bibnamefont {Wang}}, \bibinfo {author} {\bibfnamefont {A.~R.}\ \bibnamefont {Oganov}}, \bibinfo {author} {\bibfnamefont {H.}~\bibnamefont {Gou}}, \bibinfo {author} {\bibfnamefont {J.~S.}\ \bibnamefont {Smith}}, \ and\ \bibinfo {author} {\bibfnamefont {T.~A.}\ \bibnamefont {Strobel}},\ }\href@noop {} {\bibfield  {journal} {\bibinfo  {journal} {Nature Communications}\ }\textbf {\bibinfo {volume} {6}},\ \bibinfo {pages} {6974} (\bibinfo {year} {2015})}\BibitemShut {NoStop}%
\bibitem [{\citenamefont {Khandarkhaeva}\ \emph {et~al.}(2024)\citenamefont {Khandarkhaeva}, \citenamefont {Fedotenko}, \citenamefont {Aslandukova}, \citenamefont {Akbar}, \citenamefont {Bykov}, \citenamefont {Laniel}, \citenamefont {Aslandukov}, \citenamefont {Ruschewitz}, \citenamefont {Tobeck}, \citenamefont {Winkler}, \citenamefont {Chariton}, \citenamefont {Prakapenka}, \citenamefont {Glazyrin}, \citenamefont {Giacobbe}, \citenamefont {Bright}, \citenamefont {Belov}, \citenamefont {Dubrovinskaia},\ and\ \citenamefont {Dubrovinsky}}]{Khandarkhaeva2024}%
  \BibitemOpen
  \bibfield  {author} {\bibinfo {author} {\bibfnamefont {S.}~\bibnamefont {Khandarkhaeva}}, \bibinfo {author} {\bibfnamefont {T.}~\bibnamefont {Fedotenko}}, \bibinfo {author} {\bibfnamefont {A.}~\bibnamefont {Aslandukova}}, \bibinfo {author} {\bibfnamefont {F.~I.}\ \bibnamefont {Akbar}}, \bibinfo {author} {\bibfnamefont {M.}~\bibnamefont {Bykov}}, \bibinfo {author} {\bibfnamefont {D.}~\bibnamefont {Laniel}}, \bibinfo {author} {\bibfnamefont {A.}~\bibnamefont {Aslandukov}}, \bibinfo {author} {\bibfnamefont {U.}~\bibnamefont {Ruschewitz}}, \bibinfo {author} {\bibfnamefont {C.}~\bibnamefont {Tobeck}}, \bibinfo {author} {\bibfnamefont {B.}~\bibnamefont {Winkler}}, \bibinfo {author} {\bibfnamefont {S.}~\bibnamefont {Chariton}}, \bibinfo {author} {\bibfnamefont {V.}~\bibnamefont {Prakapenka}}, \bibinfo {author} {\bibfnamefont {K.}~\bibnamefont {Glazyrin}}, \bibinfo {author} {\bibfnamefont {C.}~\bibnamefont {Giacobbe}}, \bibinfo {author} {\bibfnamefont {E.~L.}\ \bibnamefont {Bright}}, \bibinfo {author}
  {\bibfnamefont {M.}~\bibnamefont {Belov}}, \bibinfo {author} {\bibfnamefont {N.}~\bibnamefont {Dubrovinskaia}}, \ and\ \bibinfo {author} {\bibfnamefont {L.}~\bibnamefont {Dubrovinsky}},\ }\href {\doibase 10.1038/s41467-024-47138-2} {\bibfield  {journal} {\bibinfo  {journal} {Nature Communications}\ }\textbf {\bibinfo {volume} {15}},\ \bibinfo {pages} {2855} (\bibinfo {year} {2024})}\BibitemShut {NoStop}%
\bibitem [{\citenamefont {Roszak}\ and\ \citenamefont {Balasubramanian}(1996)}]{roszak1996theoretical}%
  \BibitemOpen
  \bibfield  {author} {\bibinfo {author} {\bibfnamefont {S.}~\bibnamefont {Roszak}}\ and\ \bibinfo {author} {\bibfnamefont {K.}~\bibnamefont {Balasubramanian}},\ }\href@noop {} {\bibfield  {journal} {\bibinfo  {journal} {The Journal of Physical Chemistry}\ }\textbf {\bibinfo {volume} {100}},\ \bibinfo {pages} {8254} (\bibinfo {year} {1996})}\BibitemShut {NoStop}%
\bibitem [{\citenamefont {Aslandukova}\ \emph {et~al.}(2021)\citenamefont {Aslandukova}, \citenamefont {Aslandukov}, \citenamefont {Yuan}, \citenamefont {Laniel}, \citenamefont {Khandarkhaeva}, \citenamefont {Fedotenko}, \citenamefont {Steinle-Neumann}, \citenamefont {Glazyrin}, \citenamefont {Dubrovinskaia},\ and\ \citenamefont {Dubrovinsky}}]{Aslandukova2021}%
  \BibitemOpen
  \bibfield  {author} {\bibinfo {author} {\bibfnamefont {A.}~\bibnamefont {Aslandukova}}, \bibinfo {author} {\bibfnamefont {A.}~\bibnamefont {Aslandukov}}, \bibinfo {author} {\bibfnamefont {L.}~\bibnamefont {Yuan}}, \bibinfo {author} {\bibfnamefont {D.}~\bibnamefont {Laniel}}, \bibinfo {author} {\bibfnamefont {S.}~\bibnamefont {Khandarkhaeva}}, \bibinfo {author} {\bibfnamefont {T.}~\bibnamefont {Fedotenko}}, \bibinfo {author} {\bibfnamefont {G.}~\bibnamefont {Steinle-Neumann}}, \bibinfo {author} {\bibfnamefont {K.}~\bibnamefont {Glazyrin}}, \bibinfo {author} {\bibfnamefont {N.}~\bibnamefont {Dubrovinskaia}}, \ and\ \bibinfo {author} {\bibfnamefont {L.}~\bibnamefont {Dubrovinsky}},\ }\href {\doibase 10.1103/PhysRevLett.127.135501} {\bibfield  {journal} {\bibinfo  {journal} {Phys. Rev. Lett.}\ }\textbf {\bibinfo {volume} {127}},\ \bibinfo {pages} {135501} (\bibinfo {year} {2021})}\BibitemShut {NoStop}%
\bibitem [{\citenamefont {Gao}\ \emph {et~al.}(2014)\citenamefont {Gao}, \citenamefont {Jiang}, \citenamefont {Zhou},\ and\ \citenamefont {Feng}}]{gao2014stability}%
  \BibitemOpen
  \bibfield  {author} {\bibinfo {author} {\bibfnamefont {X.}~\bibnamefont {Gao}}, \bibinfo {author} {\bibfnamefont {Y.}~\bibnamefont {Jiang}}, \bibinfo {author} {\bibfnamefont {R.}~\bibnamefont {Zhou}}, \ and\ \bibinfo {author} {\bibfnamefont {J.}~\bibnamefont {Feng}},\ }\href@noop {} {\bibfield  {journal} {\bibinfo  {journal} {Journal of Alloys and Compounds}\ }\textbf {\bibinfo {volume} {587}},\ \bibinfo {pages} {819} (\bibinfo {year} {2014})}\BibitemShut {NoStop}%
\bibitem [{\citenamefont {Young~Kim}\ \emph {et~al.}(2012)\citenamefont {Young~Kim}, \citenamefont {Srepusharawoot}, \citenamefont {Pickard}, \citenamefont {Needs}, \citenamefont {Bovornratanaraks}, \citenamefont {Ahuja},\ and\ \citenamefont {Pinsook}}]{young2012phase}%
  \BibitemOpen
  \bibfield  {author} {\bibinfo {author} {\bibfnamefont {D.}~\bibnamefont {Young~Kim}}, \bibinfo {author} {\bibfnamefont {P.}~\bibnamefont {Srepusharawoot}}, \bibinfo {author} {\bibfnamefont {C.~J.}\ \bibnamefont {Pickard}}, \bibinfo {author} {\bibfnamefont {R.~J.}\ \bibnamefont {Needs}}, \bibinfo {author} {\bibfnamefont {T.}~\bibnamefont {Bovornratanaraks}}, \bibinfo {author} {\bibfnamefont {R.}~\bibnamefont {Ahuja}}, \ and\ \bibinfo {author} {\bibfnamefont {U.}~\bibnamefont {Pinsook}},\ }\href@noop {} {\bibfield  {journal} {\bibinfo  {journal} {Applied Physics Letters}\ }\textbf {\bibinfo {volume} {101}} (\bibinfo {year} {2012})}\BibitemShut {NoStop}%
\bibitem [{\citenamefont {Tsuppayakorn-Aek}\ \emph {et~al.}(2015)\citenamefont {Tsuppayakorn-Aek}, \citenamefont {Chaimayo}, \citenamefont {Pinsook},\ and\ \citenamefont {Bovornratanaraks}}]{tsuppayakorn2015existence}%
  \BibitemOpen
  \bibfield  {author} {\bibinfo {author} {\bibfnamefont {P.}~\bibnamefont {Tsuppayakorn-Aek}}, \bibinfo {author} {\bibfnamefont {W.}~\bibnamefont {Chaimayo}}, \bibinfo {author} {\bibfnamefont {U.}~\bibnamefont {Pinsook}}, \ and\ \bibinfo {author} {\bibfnamefont {T.}~\bibnamefont {Bovornratanaraks}},\ }\href@noop {} {\bibfield  {journal} {\bibinfo  {journal} {AIP Advances}\ }\textbf {\bibinfo {volume} {5}} (\bibinfo {year} {2015})}\BibitemShut {NoStop}%
\bibitem [{\citenamefont {Oganov}\ \emph {et~al.}(2011)\citenamefont {Oganov}, \citenamefont {Lyakhov},\ and\ \citenamefont {Valle}}]{Oganov2011}%
  \BibitemOpen
  \bibfield  {author} {\bibinfo {author} {\bibfnamefont {A.~R.}\ \bibnamefont {Oganov}}, \bibinfo {author} {\bibfnamefont {A.~O.}\ \bibnamefont {Lyakhov}}, \ and\ \bibinfo {author} {\bibfnamefont {M.}~\bibnamefont {Valle}},\ }\href {\doibase 10.1021/ar1001318} {\bibfield  {journal} {\bibinfo  {journal} {Accounts of Chemical Research}\ }\textbf {\bibinfo {volume} {44}},\ \bibinfo {pages} {227} (\bibinfo {year} {2011})},\ \bibinfo {note} {pMID: 21361336},\ \Eprint {http://arxiv.org/abs/https://doi.org/10.1021/ar1001318} {https://doi.org/10.1021/ar1001318} \BibitemShut {NoStop}%
\bibitem [{\citenamefont {Rybin}\ \emph {et~al.}(2021)\citenamefont {Rybin}, \citenamefont {Novoselov}, \citenamefont {Korotin}, \citenamefont {Anisimov},\ and\ \citenamefont {Oganov}}]{Rybin2021}%
  \BibitemOpen
  \bibfield  {author} {\bibinfo {author} {\bibfnamefont {N.}~\bibnamefont {Rybin}}, \bibinfo {author} {\bibfnamefont {D.~Y.}\ \bibnamefont {Novoselov}}, \bibinfo {author} {\bibfnamefont {D.~M.}\ \bibnamefont {Korotin}}, \bibinfo {author} {\bibfnamefont {V.~I.}\ \bibnamefont {Anisimov}}, \ and\ \bibinfo {author} {\bibfnamefont {A.~R.}\ \bibnamefont {Oganov}},\ }\href {\doibase 10.1039/D1CP00657F} {\bibfield  {journal} {\bibinfo  {journal} {Phys. Chem. Chem. Phys.}\ }\textbf {\bibinfo {volume} {23}},\ \bibinfo {pages} {15989} (\bibinfo {year} {2021})}\BibitemShut {NoStop}%
\bibitem [{\citenamefont {Rybin}\ \emph {et~al.}(2022)\citenamefont {Rybin}, \citenamefont {Chepkasov}, \citenamefont {Novoselov}, \citenamefont {Anisimov},\ and\ \citenamefont {Oganov}}]{Rybin2022}%
  \BibitemOpen
  \bibfield  {author} {\bibinfo {author} {\bibfnamefont {N.}~\bibnamefont {Rybin}}, \bibinfo {author} {\bibfnamefont {I.}~\bibnamefont {Chepkasov}}, \bibinfo {author} {\bibfnamefont {D.~Y.}\ \bibnamefont {Novoselov}}, \bibinfo {author} {\bibfnamefont {V.~I.}\ \bibnamefont {Anisimov}}, \ and\ \bibinfo {author} {\bibfnamefont {A.~R.}\ \bibnamefont {Oganov}},\ }\href {\doibase 10.1021/acs.jpcc.2c04785} {\bibfield  {journal} {\bibinfo  {journal} {The Journal of Physical Chemistry C}\ }\textbf {\bibinfo {volume} {126}},\ \bibinfo {pages} {15057} (\bibinfo {year} {2022})},\ \Eprint {http://arxiv.org/abs/https://doi.org/10.1021/acs.jpcc.2c04785} {https://doi.org/10.1021/acs.jpcc.2c04785} \BibitemShut {NoStop}%
\bibitem [{\citenamefont {Glass}\ \emph {et~al.}(2006)\citenamefont {Glass}, \citenamefont {Oganov},\ and\ \citenamefont {Hansen}}]{glass2006uspex}%
  \BibitemOpen
  \bibfield  {author} {\bibinfo {author} {\bibfnamefont {C.~W.}\ \bibnamefont {Glass}}, \bibinfo {author} {\bibfnamefont {A.~R.}\ \bibnamefont {Oganov}}, \ and\ \bibinfo {author} {\bibfnamefont {N.}~\bibnamefont {Hansen}},\ }\href@noop {} {\bibfield  {journal} {\bibinfo  {journal} {Computer physics communications}\ }\textbf {\bibinfo {volume} {175}},\ \bibinfo {pages} {713} (\bibinfo {year} {2006})}\BibitemShut {NoStop}%
\bibitem [{\citenamefont {Oganov}\ and\ \citenamefont {Glass}(2006)}]{Oganov2006}%
  \BibitemOpen
  \bibfield  {author} {\bibinfo {author} {\bibfnamefont {A.~R.}\ \bibnamefont {Oganov}}\ and\ \bibinfo {author} {\bibfnamefont {C.~W.}\ \bibnamefont {Glass}},\ }\href {\doibase 10.1063/1.2210932} {\bibfield  {journal} {\bibinfo  {journal} {The Journal of Chemical Physics}\ }\textbf {\bibinfo {volume} {124}},\ \bibinfo {pages} {244704} (\bibinfo {year} {2006})}\BibitemShut {NoStop}%
\bibitem [{\citenamefont {Lyakhov}\ \emph {et~al.}(2013)\citenamefont {Lyakhov}, \citenamefont {Oganov}, \citenamefont {Stokes},\ and\ \citenamefont {Zhu}}]{Lyakhov2013}%
  \BibitemOpen
  \bibfield  {author} {\bibinfo {author} {\bibfnamefont {A.~O.}\ \bibnamefont {Lyakhov}}, \bibinfo {author} {\bibfnamefont {A.~R.}\ \bibnamefont {Oganov}}, \bibinfo {author} {\bibfnamefont {H.~T.}\ \bibnamefont {Stokes}}, \ and\ \bibinfo {author} {\bibfnamefont {Q.}~\bibnamefont {Zhu}},\ }\href {\doibase 10.1016/j.cpc.2012.12.009} {\bibfield  {journal} {\bibinfo  {journal} {Computer Physics Communications}\ }\textbf {\bibinfo {volume} {184}},\ \bibinfo {pages} {1172} (\bibinfo {year} {2013})}\BibitemShut {NoStop}%
\bibitem [{\citenamefont {Jain}\ \emph {et~al.}(2015)\citenamefont {Jain}, \citenamefont {Ong}, \citenamefont {Chen}, \citenamefont {Medasani}, \citenamefont {Qu}, \citenamefont {Kocher}, \citenamefont {Brafman}, \citenamefont {Petretto}, \citenamefont {Rignanese}, \citenamefont {Hautier}, \citenamefont {Gunter},\ and\ \citenamefont {Persson}}]{Jain2015}%
  \BibitemOpen
  \bibfield  {author} {\bibinfo {author} {\bibfnamefont {A.}~\bibnamefont {Jain}}, \bibinfo {author} {\bibfnamefont {S.~P.}\ \bibnamefont {Ong}}, \bibinfo {author} {\bibfnamefont {W.}~\bibnamefont {Chen}}, \bibinfo {author} {\bibfnamefont {B.}~\bibnamefont {Medasani}}, \bibinfo {author} {\bibfnamefont {X.}~\bibnamefont {Qu}}, \bibinfo {author} {\bibfnamefont {M.}~\bibnamefont {Kocher}}, \bibinfo {author} {\bibfnamefont {M.}~\bibnamefont {Brafman}}, \bibinfo {author} {\bibfnamefont {G.}~\bibnamefont {Petretto}}, \bibinfo {author} {\bibfnamefont {G.-M.}\ \bibnamefont {Rignanese}}, \bibinfo {author} {\bibfnamefont {G.}~\bibnamefont {Hautier}}, \bibinfo {author} {\bibfnamefont {D.}~\bibnamefont {Gunter}}, \ and\ \bibinfo {author} {\bibfnamefont {K.~A.}\ \bibnamefont {Persson}},\ }\href {\doibase https://doi.org/10.1002/cpe.3505} {\bibfield  {journal} {\bibinfo  {journal} {Concurrency and Computation: Practice and Experience}\ }\textbf {\bibinfo {volume} {27}},\ \bibinfo {pages} {5037} (\bibinfo {year} {2015})},\
  \Eprint {http://arxiv.org/abs/https://onlinelibrary.wiley.com/doi/pdf/10.1002/cpe.3505} {https://onlinelibrary.wiley.com/doi/pdf/10.1002/cpe.3505} \BibitemShut {NoStop}%
\bibitem [{\citenamefont {Deaven}\ and\ \citenamefont {Ho}(1995)}]{PhysRevLett.75.288}%
  \BibitemOpen
  \bibfield  {author} {\bibinfo {author} {\bibfnamefont {D.~M.}\ \bibnamefont {Deaven}}\ and\ \bibinfo {author} {\bibfnamefont {K.~M.}\ \bibnamefont {Ho}},\ }\href {\doibase 10.1103/PhysRevLett.75.288} {\bibfield  {journal} {\bibinfo  {journal} {Phys. Rev. Lett.}\ }\textbf {\bibinfo {volume} {75}},\ \bibinfo {pages} {288} (\bibinfo {year} {1995})}\BibitemShut {NoStop}%
\bibitem [{\citenamefont {Perdew}\ \emph {et~al.}(1996)\citenamefont {Perdew}, \citenamefont {Burke},\ and\ \citenamefont {Ernzerhof}}]{Perdew1996a}%
  \BibitemOpen
  \bibfield  {author} {\bibinfo {author} {\bibfnamefont {J.~P.}\ \bibnamefont {Perdew}}, \bibinfo {author} {\bibfnamefont {K.}~\bibnamefont {Burke}}, \ and\ \bibinfo {author} {\bibfnamefont {M.}~\bibnamefont {Ernzerhof}},\ }\href {\doibase 10.1103/PhysRevLett.77.3865} {\bibfield  {journal} {\bibinfo  {journal} {Physical Review Letters}\ }\textbf {\bibinfo {volume} {77}},\ \bibinfo {pages} {3865} (\bibinfo {year} {1996})}\BibitemShut {NoStop}%
\bibitem [{\citenamefont {Giannozzi}\ \emph {et~al.}(2009)\citenamefont {Giannozzi}, \citenamefont {Baroni}, \citenamefont {Bonini}, \citenamefont {Calandra}, \citenamefont {Car}, \citenamefont {Cavazzoni}, \citenamefont {Ceresoli}, \citenamefont {Chiarotti}, \citenamefont {Cococcioni}, \citenamefont {Dabo} \emph {et~al.}}]{giannozzi2009quantum}%
  \BibitemOpen
  \bibfield  {author} {\bibinfo {author} {\bibfnamefont {P.}~\bibnamefont {Giannozzi}}, \bibinfo {author} {\bibfnamefont {S.}~\bibnamefont {Baroni}}, \bibinfo {author} {\bibfnamefont {N.}~\bibnamefont {Bonini}}, \bibinfo {author} {\bibfnamefont {M.}~\bibnamefont {Calandra}}, \bibinfo {author} {\bibfnamefont {R.}~\bibnamefont {Car}}, \bibinfo {author} {\bibfnamefont {C.}~\bibnamefont {Cavazzoni}}, \bibinfo {author} {\bibfnamefont {D.}~\bibnamefont {Ceresoli}}, \bibinfo {author} {\bibfnamefont {G.~L.}\ \bibnamefont {Chiarotti}}, \bibinfo {author} {\bibfnamefont {M.}~\bibnamefont {Cococcioni}}, \bibinfo {author} {\bibfnamefont {I.}~\bibnamefont {Dabo}},  \emph {et~al.},\ }\href@noop {} {\bibfield  {journal} {\bibinfo  {journal} {Journal of physics: Condensed matter}\ }\textbf {\bibinfo {volume} {21}},\ \bibinfo {pages} {395502} (\bibinfo {year} {2009})}\BibitemShut {NoStop}%
\bibitem [{\citenamefont {Giannozzi}\ \emph {et~al.}(2017)\citenamefont {Giannozzi} \emph {et~al.}}]{Giannozzi_2017}%
  \BibitemOpen
  \bibfield  {author} {\bibinfo {author} {\bibfnamefont {P.}~\bibnamefont {Giannozzi}} \emph {et~al.},\ }\href {\doibase 10.1088/1361-648X/aa8f79} {\bibfield  {journal} {\bibinfo  {journal} {Journal of Physics: Condensed Matter}\ }\textbf {\bibinfo {volume} {29}},\ \bibinfo {pages} {465901} (\bibinfo {year} {2017})}\BibitemShut {NoStop}%
\bibitem [{\citenamefont {Kresse}\ and\ \citenamefont {Furthm{\"{u}}ller}(1996)}]{VASP}%
  \BibitemOpen
  \bibfield  {author} {\bibinfo {author} {\bibfnamefont {G.}~\bibnamefont {Kresse}}\ and\ \bibinfo {author} {\bibfnamefont {J.}~\bibnamefont {Furthm{\"{u}}ller}},\ }\href {\doibase 10.1103/PhysRevB.54.11169} {\bibfield  {journal} {\bibinfo  {journal} {Physical Review B}\ }\textbf {\bibinfo {volume} {54}},\ \bibinfo {pages} {11169} (\bibinfo {year} {1996})}\BibitemShut {NoStop}%
\bibitem [{\citenamefont {Bl{\"{o}}chl}(1994)}]{Blochl1994}%
  \BibitemOpen
  \bibfield  {author} {\bibinfo {author} {\bibfnamefont {P.~E.}\ \bibnamefont {Bl{\"{o}}chl}},\ }\href {\doibase 10.1103/PhysRevB.50.17953} {\bibfield  {journal} {\bibinfo  {journal} {Physical Review B}\ }\textbf {\bibinfo {volume} {50}},\ \bibinfo {pages} {17953} (\bibinfo {year} {1994})}\BibitemShut {NoStop}%
\bibitem [{\citenamefont {Fredericks}\ \emph {et~al.}(2021)\citenamefont {Fredericks}, \citenamefont {Parrish}, \citenamefont {Sayre},\ and\ \citenamefont {Zhu}}]{fredericks2021pyxtal}%
  \BibitemOpen
  \bibfield  {author} {\bibinfo {author} {\bibfnamefont {S.}~\bibnamefont {Fredericks}}, \bibinfo {author} {\bibfnamefont {K.}~\bibnamefont {Parrish}}, \bibinfo {author} {\bibfnamefont {D.}~\bibnamefont {Sayre}}, \ and\ \bibinfo {author} {\bibfnamefont {Q.}~\bibnamefont {Zhu}},\ }\href@noop {} {\bibfield  {journal} {\bibinfo  {journal} {Computer Physics Communications}\ }\textbf {\bibinfo {volume} {261}},\ \bibinfo {pages} {107810} (\bibinfo {year} {2021})}\BibitemShut {NoStop}%
\bibitem [{\citenamefont {Bushlanov}\ \emph {et~al.}(2019)\citenamefont {Bushlanov}, \citenamefont {Blatov},\ and\ \citenamefont {Oganov}}]{Bushlanov2019}%
  \BibitemOpen
  \bibfield  {author} {\bibinfo {author} {\bibfnamefont {P.~V.}\ \bibnamefont {Bushlanov}}, \bibinfo {author} {\bibfnamefont {V.~A.}\ \bibnamefont {Blatov}}, \ and\ \bibinfo {author} {\bibfnamefont {A.~R.}\ \bibnamefont {Oganov}},\ }\href {\doibase 10.1016/j.cpc.2018.09.016} {\bibfield  {journal} {\bibinfo  {journal} {Computer Physics Communications}\ }\textbf {\bibinfo {volume} {236}},\ \bibinfo {pages} {1} (\bibinfo {year} {2019})}\BibitemShut {NoStop}%
\bibitem [{\citenamefont {Parlinski}\ \emph {et~al.}(1997)\citenamefont {Parlinski}, \citenamefont {Li},\ and\ \citenamefont {Kawazoe}}]{Parlinski1997}%
  \BibitemOpen
  \bibfield  {author} {\bibinfo {author} {\bibfnamefont {K.}~\bibnamefont {Parlinski}}, \bibinfo {author} {\bibfnamefont {Z.~Q.}\ \bibnamefont {Li}}, \ and\ \bibinfo {author} {\bibfnamefont {Y.}~\bibnamefont {Kawazoe}},\ }\href {\doibase 10.1103/PhysRevLett.78.4063} {\bibfield  {journal} {\bibinfo  {journal} {Phys. Rev. Lett.}\ }\textbf {\bibinfo {volume} {78}},\ \bibinfo {pages} {4063} (\bibinfo {year} {1997})}\BibitemShut {NoStop}%
\bibitem [{\citenamefont {Togo}\ and\ \citenamefont {Tanaka}(2015)}]{Togo2015}%
  \BibitemOpen
  \bibfield  {author} {\bibinfo {author} {\bibfnamefont {A.}~\bibnamefont {Togo}}\ and\ \bibinfo {author} {\bibfnamefont {I.}~\bibnamefont {Tanaka}},\ }\href {\doibase 10.1016/j.scriptamat.2015.07.021} {\bibfield  {journal} {\bibinfo  {journal} {Scripta Materialia}\ }\textbf {\bibinfo {volume} {108}},\ \bibinfo {pages} {1} (\bibinfo {year} {2015})}\BibitemShut {NoStop}%
\bibitem [{\citenamefont {Momma}\ and\ \citenamefont {Izumi}(2011)}]{momma2011vesta}%
  \BibitemOpen
  \bibfield  {author} {\bibinfo {author} {\bibfnamefont {K.}~\bibnamefont {Momma}}\ and\ \bibinfo {author} {\bibfnamefont {F.}~\bibnamefont {Izumi}},\ }\href@noop {} {\bibfield  {journal} {\bibinfo  {journal} {Journal of applied crystallography}\ }\textbf {\bibinfo {volume} {44}},\ \bibinfo {pages} {1272} (\bibinfo {year} {2011})}\BibitemShut {NoStop}%
\bibitem [{\citenamefont {Henkelman}\ \emph {et~al.}(2006)\citenamefont {Henkelman}, \citenamefont {Arnaldsson},\ and\ \citenamefont {J{\'o}nsson}}]{henkelman2006fast}%
  \BibitemOpen
  \bibfield  {author} {\bibinfo {author} {\bibfnamefont {G.}~\bibnamefont {Henkelman}}, \bibinfo {author} {\bibfnamefont {A.}~\bibnamefont {Arnaldsson}}, \ and\ \bibinfo {author} {\bibfnamefont {H.}~\bibnamefont {J{\'o}nsson}},\ }\href@noop {} {\bibfield  {journal} {\bibinfo  {journal} {Computational Materials Science}\ }\textbf {\bibinfo {volume} {36}},\ \bibinfo {pages} {354} (\bibinfo {year} {2006})}\BibitemShut {NoStop}%
\bibitem [{\citenamefont {Yu}\ and\ \citenamefont {Trinkle}(2011)}]{yu2011accurate}%
  \BibitemOpen
  \bibfield  {author} {\bibinfo {author} {\bibfnamefont {M.}~\bibnamefont {Yu}}\ and\ \bibinfo {author} {\bibfnamefont {D.~R.}\ \bibnamefont {Trinkle}},\ }\href@noop {} {\bibfield  {journal} {\bibinfo  {journal} {The Journal of chemical physics}\ }\textbf {\bibinfo {volume} {134}} (\bibinfo {year} {2011})}\BibitemShut {NoStop}%
\bibitem [{\citenamefont {Sanville}\ \emph {et~al.}(2007)\citenamefont {Sanville}, \citenamefont {Kenny}, \citenamefont {Smith},\ and\ \citenamefont {Henkelman}}]{sanville2007improved}%
  \BibitemOpen
  \bibfield  {author} {\bibinfo {author} {\bibfnamefont {E.}~\bibnamefont {Sanville}}, \bibinfo {author} {\bibfnamefont {S.~D.}\ \bibnamefont {Kenny}}, \bibinfo {author} {\bibfnamefont {R.}~\bibnamefont {Smith}}, \ and\ \bibinfo {author} {\bibfnamefont {G.}~\bibnamefont {Henkelman}},\ }\href@noop {} {\bibfield  {journal} {\bibinfo  {journal} {Journal of computational chemistry}\ }\textbf {\bibinfo {volume} {28}},\ \bibinfo {pages} {899} (\bibinfo {year} {2007})}\BibitemShut {NoStop}%
\bibitem [{\citenamefont {Tang}\ \emph {et~al.}(2009)\citenamefont {Tang}, \citenamefont {Sanville},\ and\ \citenamefont {Henkelman}}]{tang2009grid}%
  \BibitemOpen
  \bibfield  {author} {\bibinfo {author} {\bibfnamefont {W.}~\bibnamefont {Tang}}, \bibinfo {author} {\bibfnamefont {E.}~\bibnamefont {Sanville}}, \ and\ \bibinfo {author} {\bibfnamefont {G.}~\bibnamefont {Henkelman}},\ }\href@noop {} {\bibfield  {journal} {\bibinfo  {journal} {Journal of Physics: Condensed Matter}\ }\textbf {\bibinfo {volume} {21}},\ \bibinfo {pages} {084204} (\bibinfo {year} {2009})}\BibitemShut {NoStop}%
\bibitem [{\citenamefont {Kurakevych}\ \emph {et~al.}(2013)\citenamefont {Kurakevych}, \citenamefont {Strobel}, \citenamefont {Kim},\ and\ \citenamefont {Cody}}]{kurakevych2013synthesis}%
  \BibitemOpen
  \bibfield  {author} {\bibinfo {author} {\bibfnamefont {O.~O.}\ \bibnamefont {Kurakevych}}, \bibinfo {author} {\bibfnamefont {T.~A.}\ \bibnamefont {Strobel}}, \bibinfo {author} {\bibfnamefont {D.~Y.}\ \bibnamefont {Kim}}, \ and\ \bibinfo {author} {\bibfnamefont {G.~D.}\ \bibnamefont {Cody}},\ }\href@noop {} {\bibfield  {journal} {\bibinfo  {journal} {Angewandte Chemie International Edition}\ }\textbf {\bibinfo {volume} {52}},\ \bibinfo {pages} {8930} (\bibinfo {year} {2013})}\BibitemShut {NoStop}%
\bibitem [{\citenamefont {Zachariasen}(1948)}]{zachariasen1948crystal}%
  \BibitemOpen
  \bibfield  {author} {\bibinfo {author} {\bibfnamefont {W.}~\bibnamefont {Zachariasen}},\ }\href@noop {} {\bibfield  {journal} {\bibinfo  {journal} {Acta Crystallographica}\ }\textbf {\bibinfo {volume} {1}},\ \bibinfo {pages} {265} (\bibinfo {year} {1948})}\BibitemShut {NoStop}%
\bibitem [{\citenamefont {Riabov}\ \emph {et~al.}(1999)\citenamefont {Riabov}, \citenamefont {Yartys}, \citenamefont {Hauback}, \citenamefont {Guegan}, \citenamefont {Wiesinger},\ and\ \citenamefont {Harris}}]{riabov1999hydrogenation}%
  \BibitemOpen
  \bibfield  {author} {\bibinfo {author} {\bibfnamefont {A.}~\bibnamefont {Riabov}}, \bibinfo {author} {\bibfnamefont {V.}~\bibnamefont {Yartys}}, \bibinfo {author} {\bibfnamefont {B.}~\bibnamefont {Hauback}}, \bibinfo {author} {\bibfnamefont {P.}~\bibnamefont {Guegan}}, \bibinfo {author} {\bibfnamefont {G.}~\bibnamefont {Wiesinger}}, \ and\ \bibinfo {author} {\bibfnamefont {I.}~\bibnamefont {Harris}},\ }\href@noop {} {\bibfield  {journal} {\bibinfo  {journal} {Journal of alloys and compounds}\ }\textbf {\bibinfo {volume} {293}},\ \bibinfo {pages} {93} (\bibinfo {year} {1999})}\BibitemShut {NoStop}%
\bibitem [{\citenamefont {Chai}\ and\ \citenamefont {Corbett}(2011)}]{chai2011two}%
  \BibitemOpen
  \bibfield  {author} {\bibinfo {author} {\bibfnamefont {P.}~\bibnamefont {Chai}}\ and\ \bibinfo {author} {\bibfnamefont {J.~D.}\ \bibnamefont {Corbett}},\ }\href@noop {} {\bibfield  {journal} {\bibinfo  {journal} {Acta Crystallographica Section C: Crystal Structure Communications}\ }\textbf {\bibinfo {volume} {67}},\ \bibinfo {pages} {i53} (\bibinfo {year} {2011})}\BibitemShut {NoStop}%
\bibitem [{\citenamefont {Almenningen}\ \emph {et~al.}(1959)\citenamefont {Almenningen}, \citenamefont {Bastiansen},\ and\ \citenamefont {Traetteberg}}]{almenningen1959electron}%
  \BibitemOpen
  \bibfield  {author} {\bibinfo {author} {\bibfnamefont {A.}~\bibnamefont {Almenningen}}, \bibinfo {author} {\bibfnamefont {O.}~\bibnamefont {Bastiansen}}, \ and\ \bibinfo {author} {\bibfnamefont {M.}~\bibnamefont {Traetteberg}},\ }\href@noop {} {\bibfield  {journal} {\bibinfo  {journal} {Acta Chemica Scandinavica}\ }\textbf {\bibinfo {volume} {13}},\ \bibinfo {pages} {1699} (\bibinfo {year} {1959})}\BibitemShut {NoStop}%
\bibitem [{\citenamefont {Akbar}\ \emph {et~al.}(2024{\natexlab{b}})\citenamefont {Akbar}, \citenamefont {Aslandukova}, \citenamefont {Yin}, \citenamefont {Aslandukov}, \citenamefont {Laniel}, \citenamefont {Bykova}, \citenamefont {Bykov}, \citenamefont {Bright}, \citenamefont {Wright}, \citenamefont {Comboni} \emph {et~al.}}]{akbar2024high}%
  \BibitemOpen
  \bibfield  {author} {\bibinfo {author} {\bibfnamefont {F.~I.}\ \bibnamefont {Akbar}}, \bibinfo {author} {\bibfnamefont {A.}~\bibnamefont {Aslandukova}}, \bibinfo {author} {\bibfnamefont {Y.}~\bibnamefont {Yin}}, \bibinfo {author} {\bibfnamefont {A.}~\bibnamefont {Aslandukov}}, \bibinfo {author} {\bibfnamefont {D.}~\bibnamefont {Laniel}}, \bibinfo {author} {\bibfnamefont {E.}~\bibnamefont {Bykova}}, \bibinfo {author} {\bibfnamefont {M.}~\bibnamefont {Bykov}}, \bibinfo {author} {\bibfnamefont {E.~L.}\ \bibnamefont {Bright}}, \bibinfo {author} {\bibfnamefont {J.}~\bibnamefont {Wright}}, \bibinfo {author} {\bibfnamefont {D.}~\bibnamefont {Comboni}},  \emph {et~al.},\ }\href@noop {} {\bibfield  {journal} {\bibinfo  {journal} {Carbon}\ ,\ \bibinfo {pages} {119374}} (\bibinfo {year} {2024}{\natexlab{b}})}\BibitemShut {NoStop}%
\end{thebibliography}%
\end{document}


\begin{figure*}[h!]
\begin{minipage}[h]{1\linewidth}
\center{\includegraphics[width=1\linewidth]{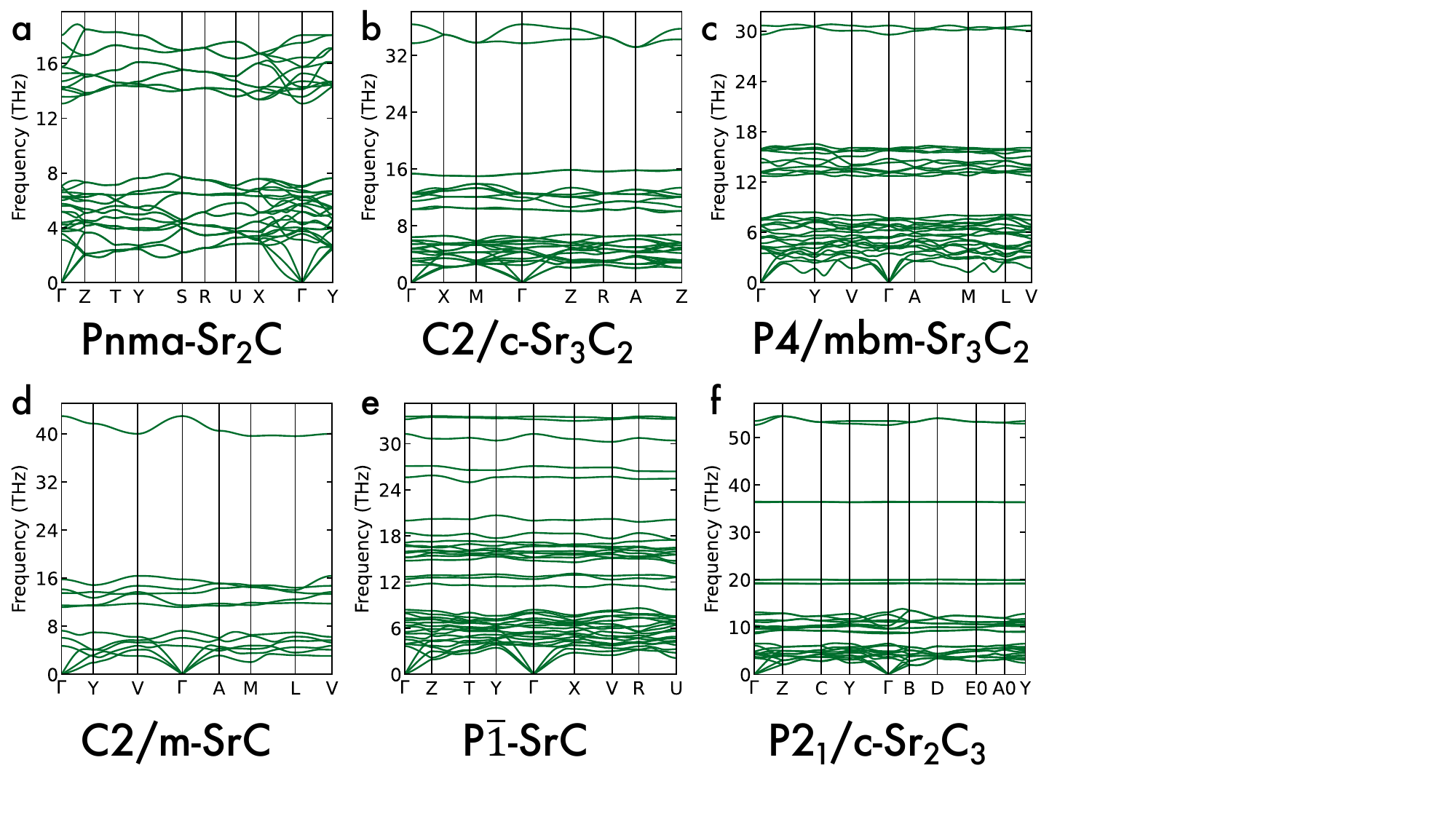}}
\end{minipage}
\hfill
\begin{minipage}[h]{1\linewidth}
\center{\includegraphics[width=1\linewidth]{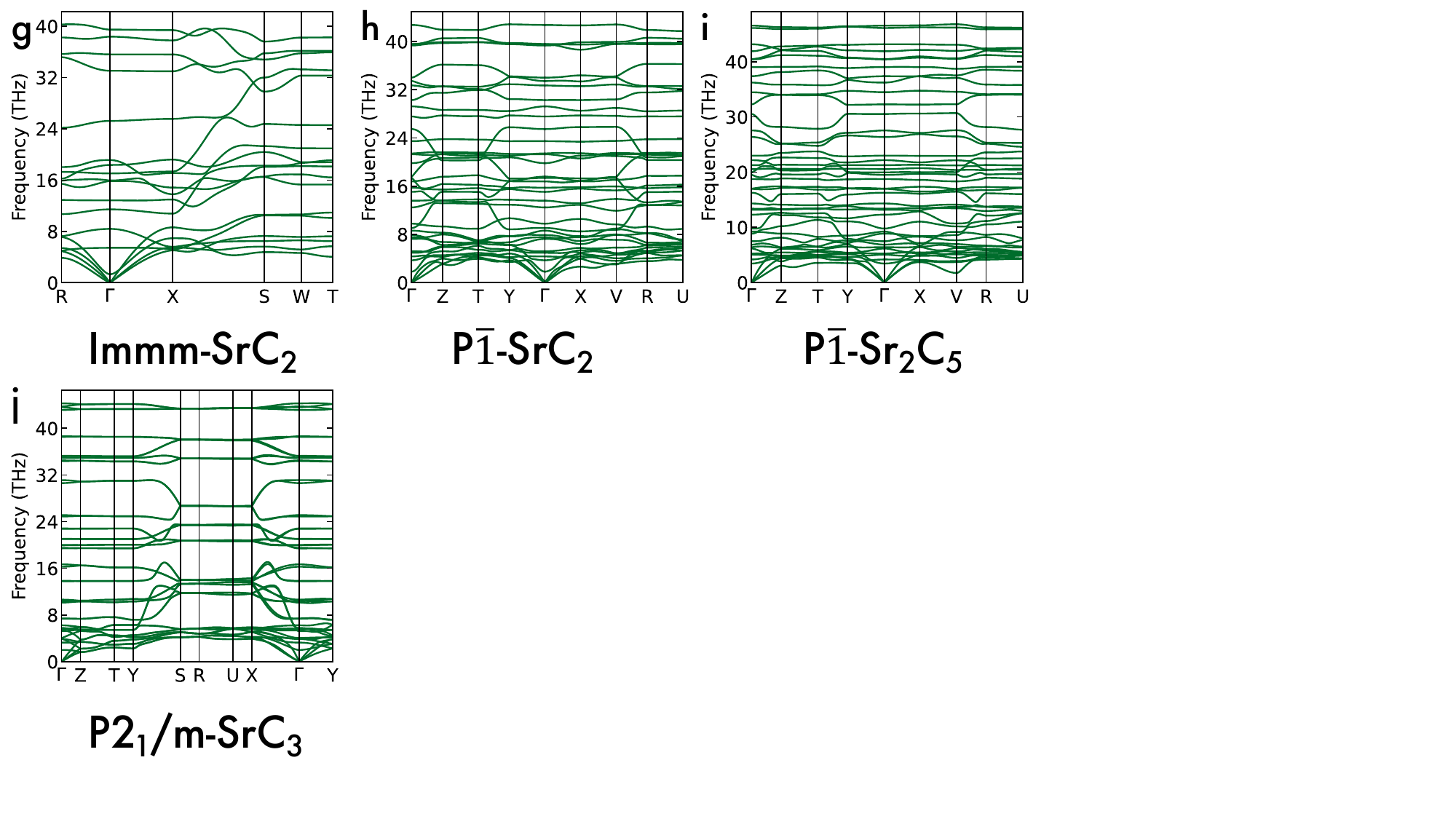}} 
\end{minipage}
\caption {Phonon dispersion curves for discovered Sr-C compounds.} 
\label{fig:ph}
\end{figure*}

\begin{figure*}[h!]
\begin{minipage}[h]{1\linewidth}
\center{\includegraphics[width=1\linewidth]{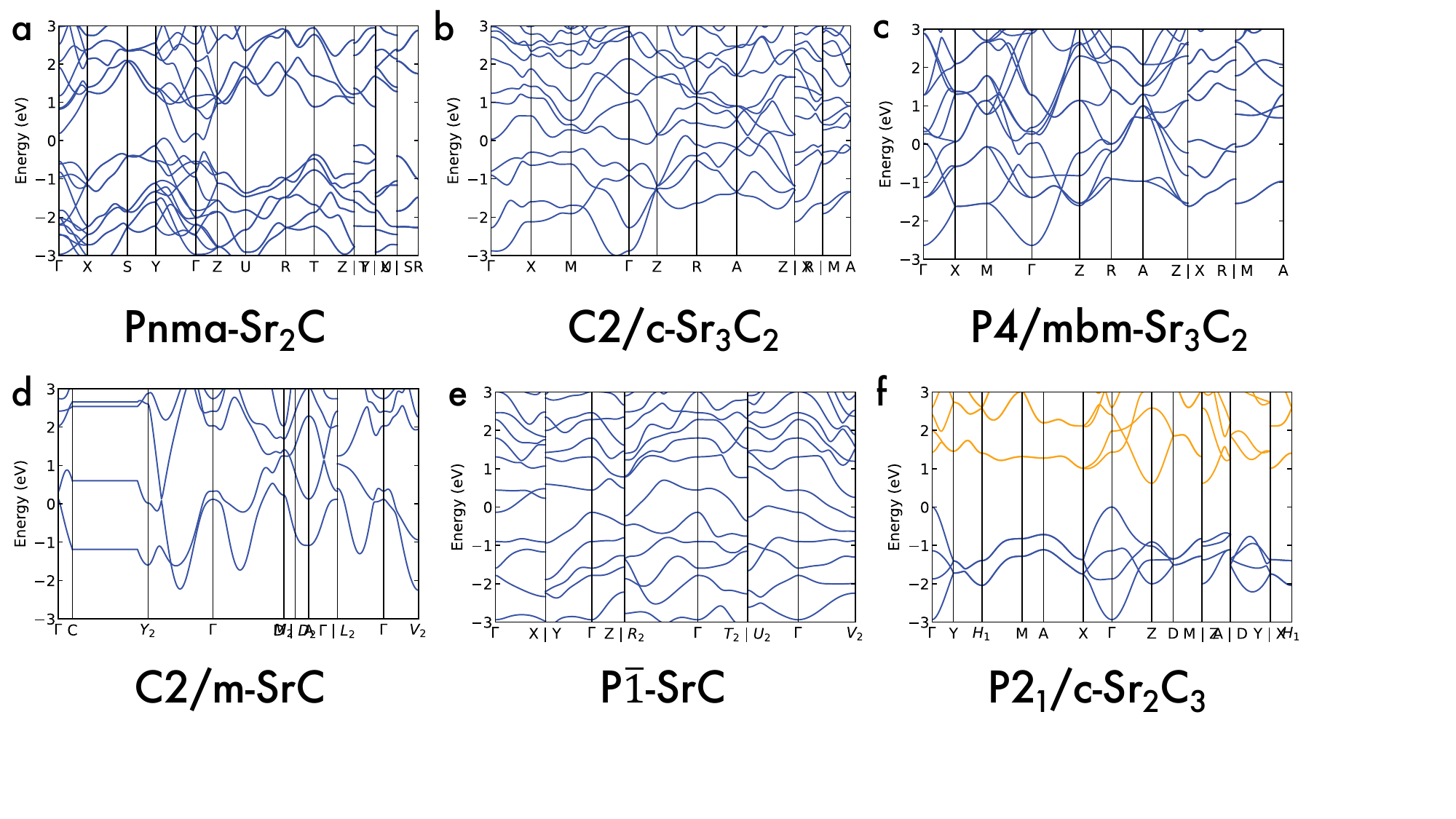}}
\end{minipage}
\hfill
\begin{minipage}[h]{1\linewidth}
\center{\includegraphics[width=1\linewidth]{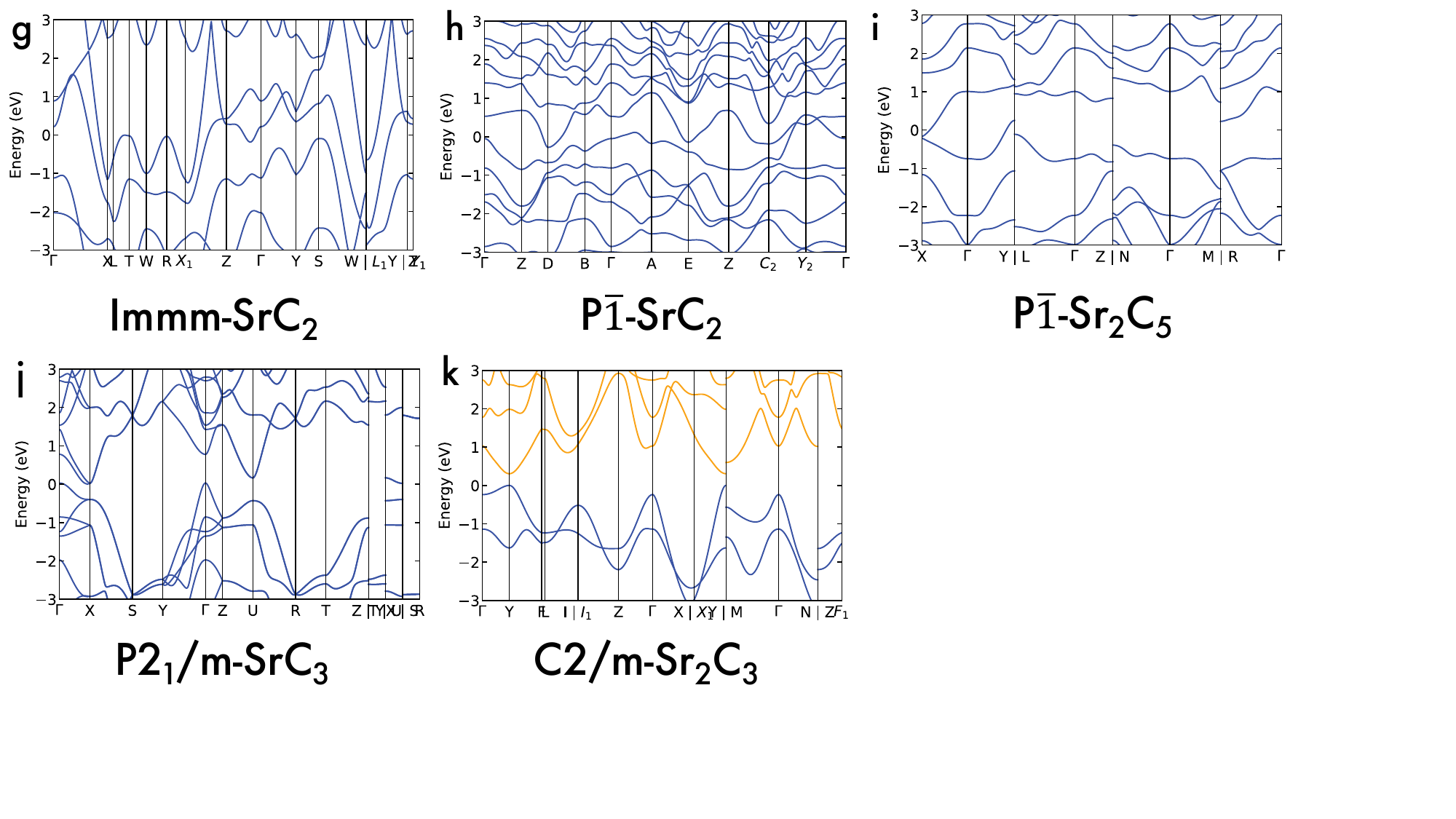}}
\end{minipage}
\caption {Band structures for discovered Sr-C compounds.} 
\label{fig:ph}
\end{figure*}

\begin{figure}[h!]
	\centering
	\begin{minipage}[h]{1\linewidth}
		\center{\includegraphics[trim={0cm 0cm 0cm 0cm},clip, width=0.75\linewidth]{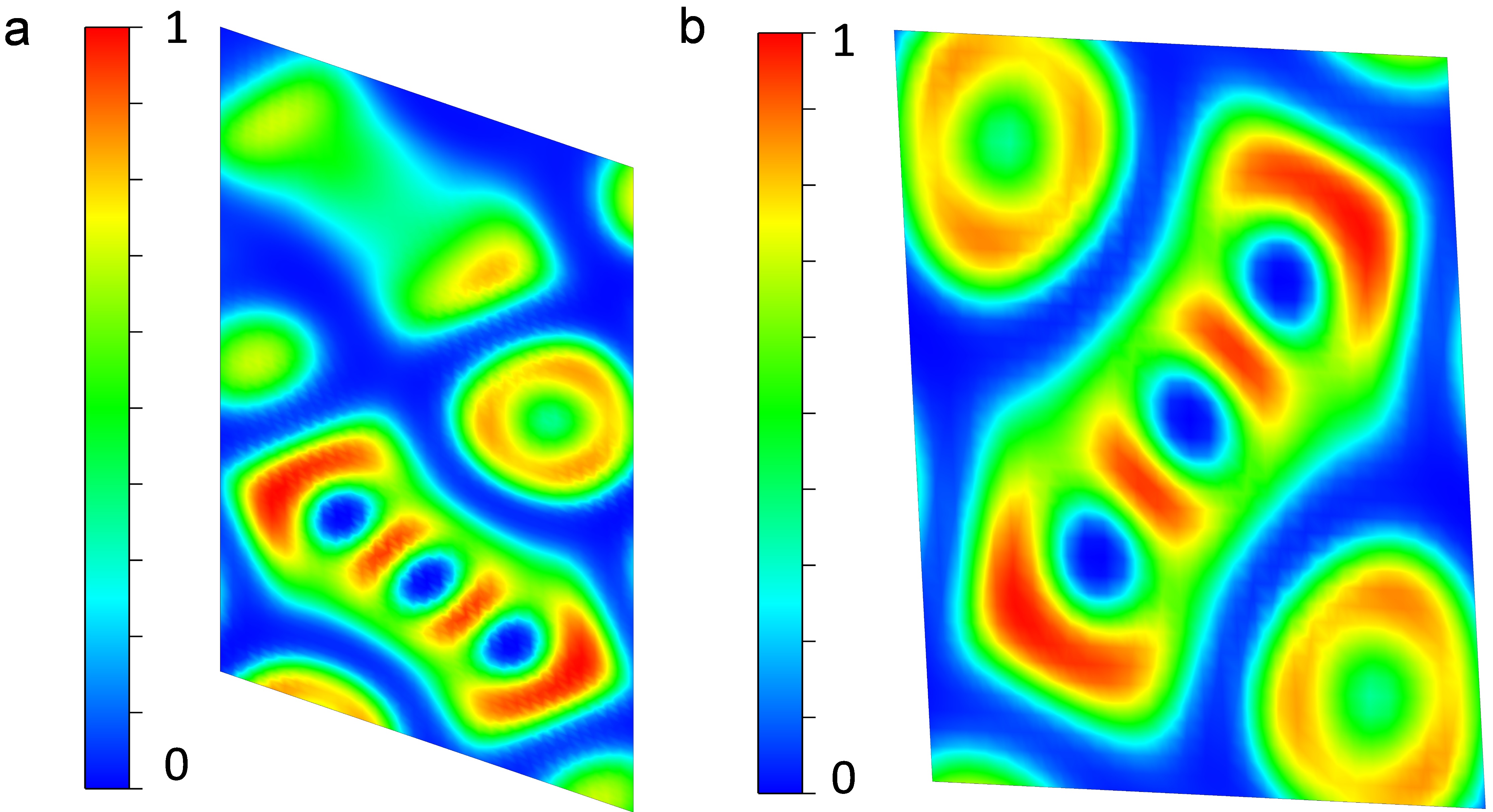}}
	\end{minipage}
    \caption{2D electron localizatio function of (a) $C2/m$-Sr$_2$C$_3$ structure at 20 GPa and (b) $P2_{1}/c$-Sr$_2$C$_3$ structure at 30 GPa, showing the covalent bonding between the carbon atoms and ionic interaction between Sr and C atoms.}
      \label{s2c3:src}
\end{figure}

\begin{figure}[h!]
	\centering
	\begin{minipage}[h]{1\linewidth}
		\center{\includegraphics[trim={0cm 0cm 0cm 0cm},clip, width=0.75\linewidth]{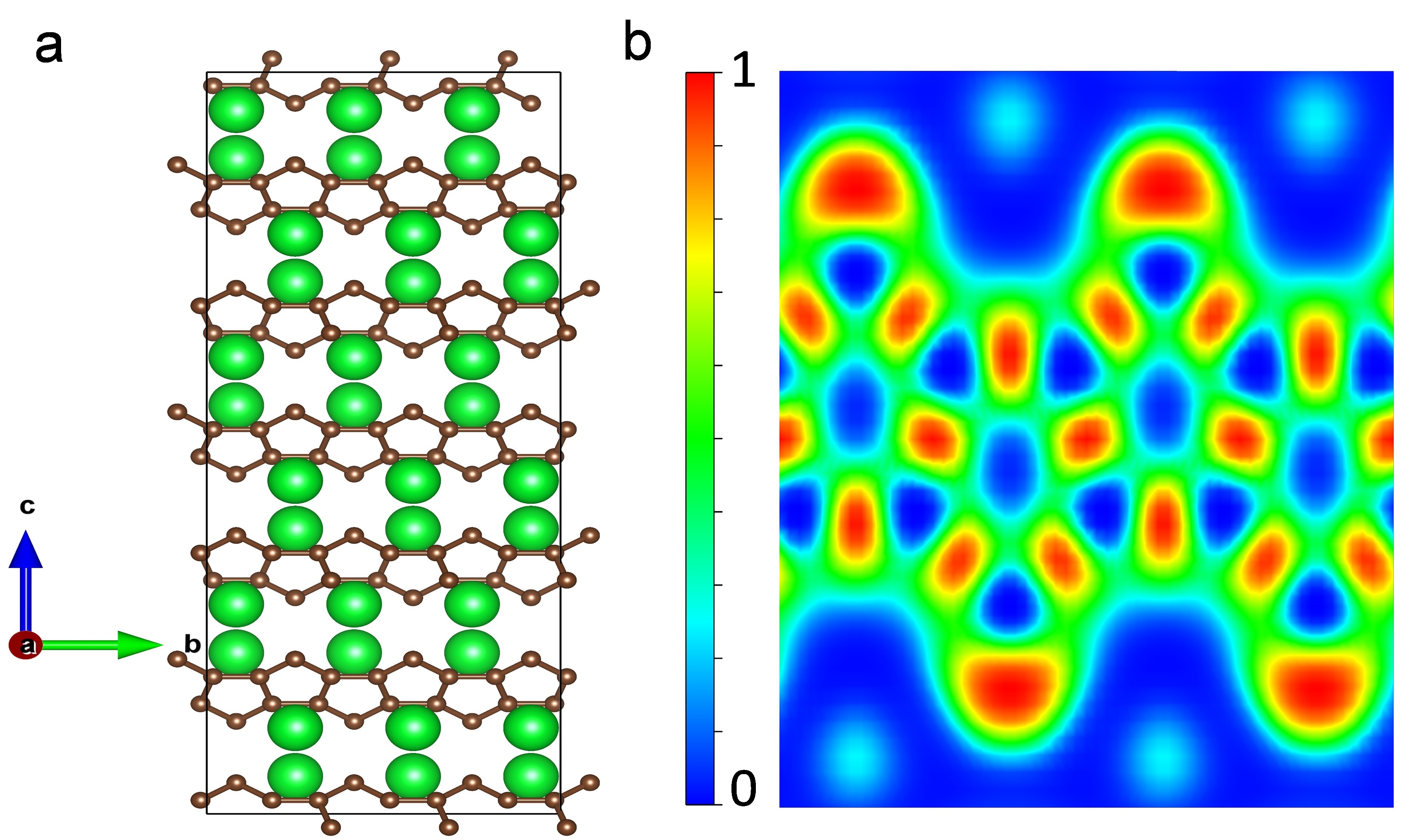}}
	\end{minipage}
    \caption{(a) The crystal structure of SrC$_3$ viewd along a-direction showing the infinite pentagonal ribbons (b) The 2D electron localization function containing the carbon ribbons}
\end{figure}

\begin{figure}[h!]
	\centering
	\begin{minipage}[h]{1\linewidth}
		\center{\includegraphics[trim={0cm 0cm 0cm 0cm},clip, width=0.75\linewidth]{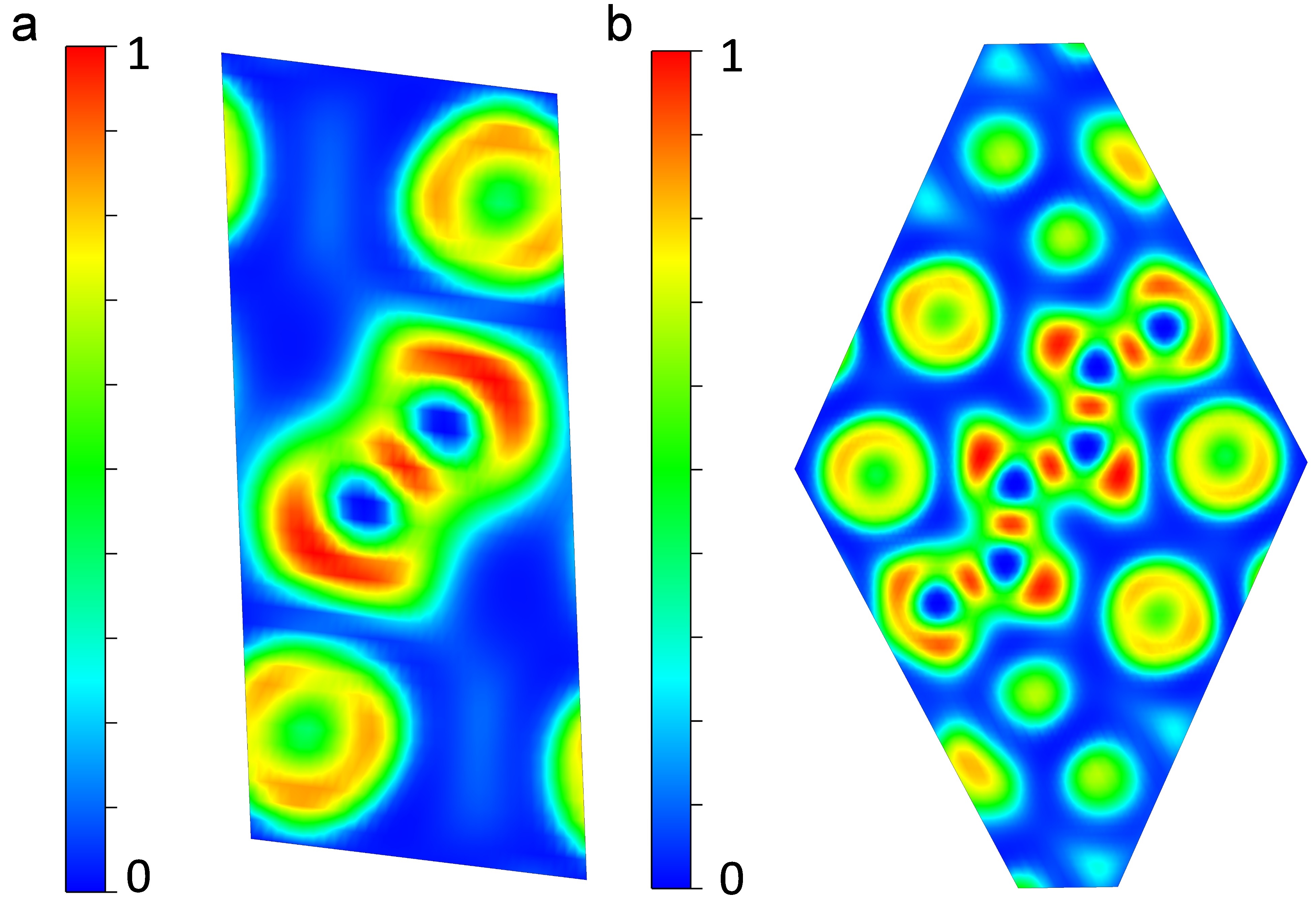}}
	\end{minipage}
    \caption{The 2D Electron Localization Function (ELF) shows (a) electron density localization around the C-dimer in $C2/m$-SrC at 30 GPa, and (b) around the 6-atom carbon chain in $P\bar{1}$-SrC at 50 GPa. Both ELFs indicate ionic Sr–C interactions and covalent bonding between carbon atoms.}
\end{figure}

\begin{figure}[h!]
	\centering
	\begin{minipage}[h]{1\linewidth}
		\center{\includegraphics[trim={0cm 0cm 0cm 0cm},clip, width=0.75\linewidth]{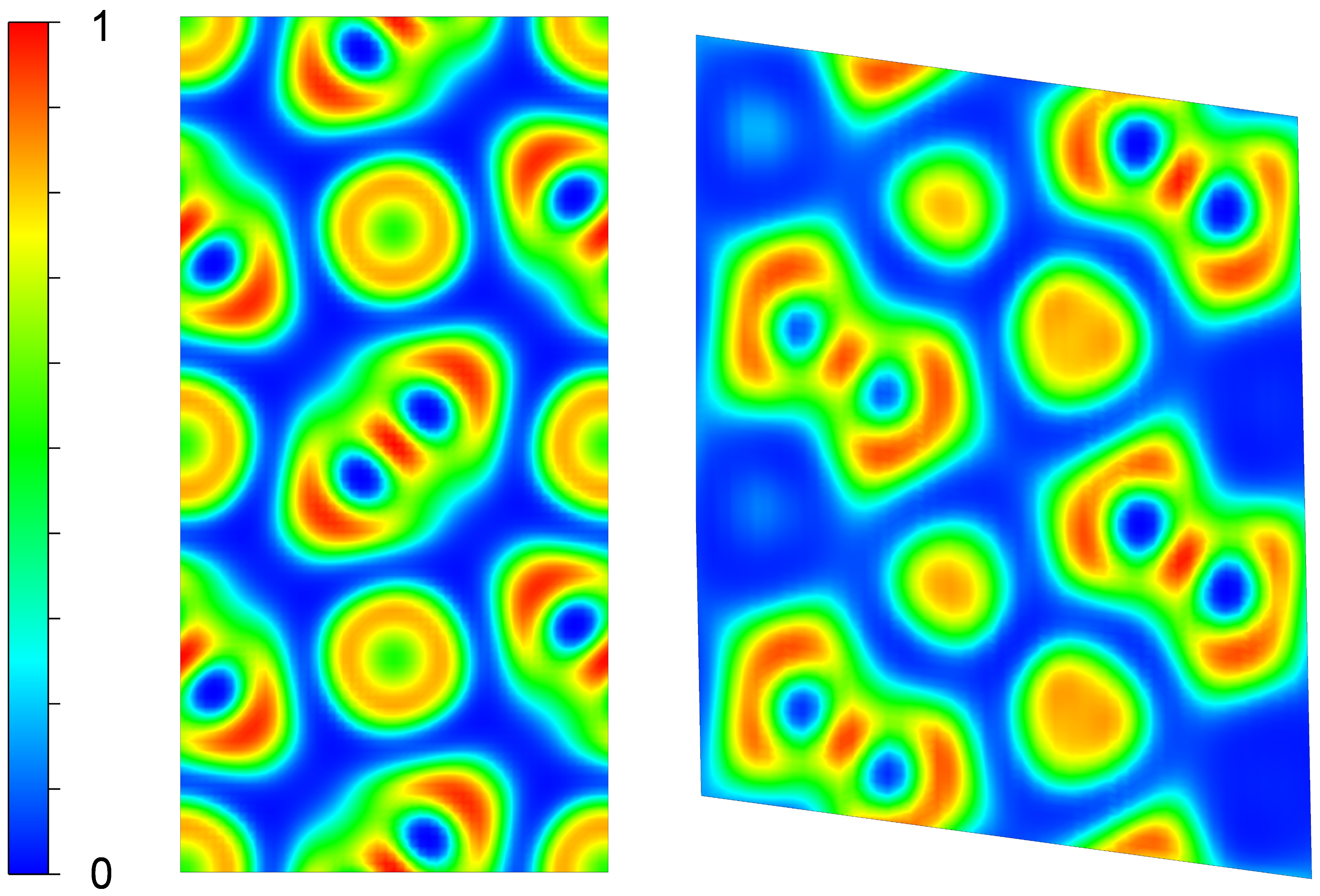}}
	\end{minipage}
    \caption{The 2D Electron Localization Function (ELF) containing the C$_2$ dimers in $P4/mbm$-Sr$_3$C$_2$ (left) at 30GPa and in $C2/c$-Sr$_3$C$_2$ (right) at 50 GPa.}
\end{figure}

\begin{figure*}[hbt!]
	\centering
	\makebox[0pt][c]{\includegraphics[width=1.2\linewidth]{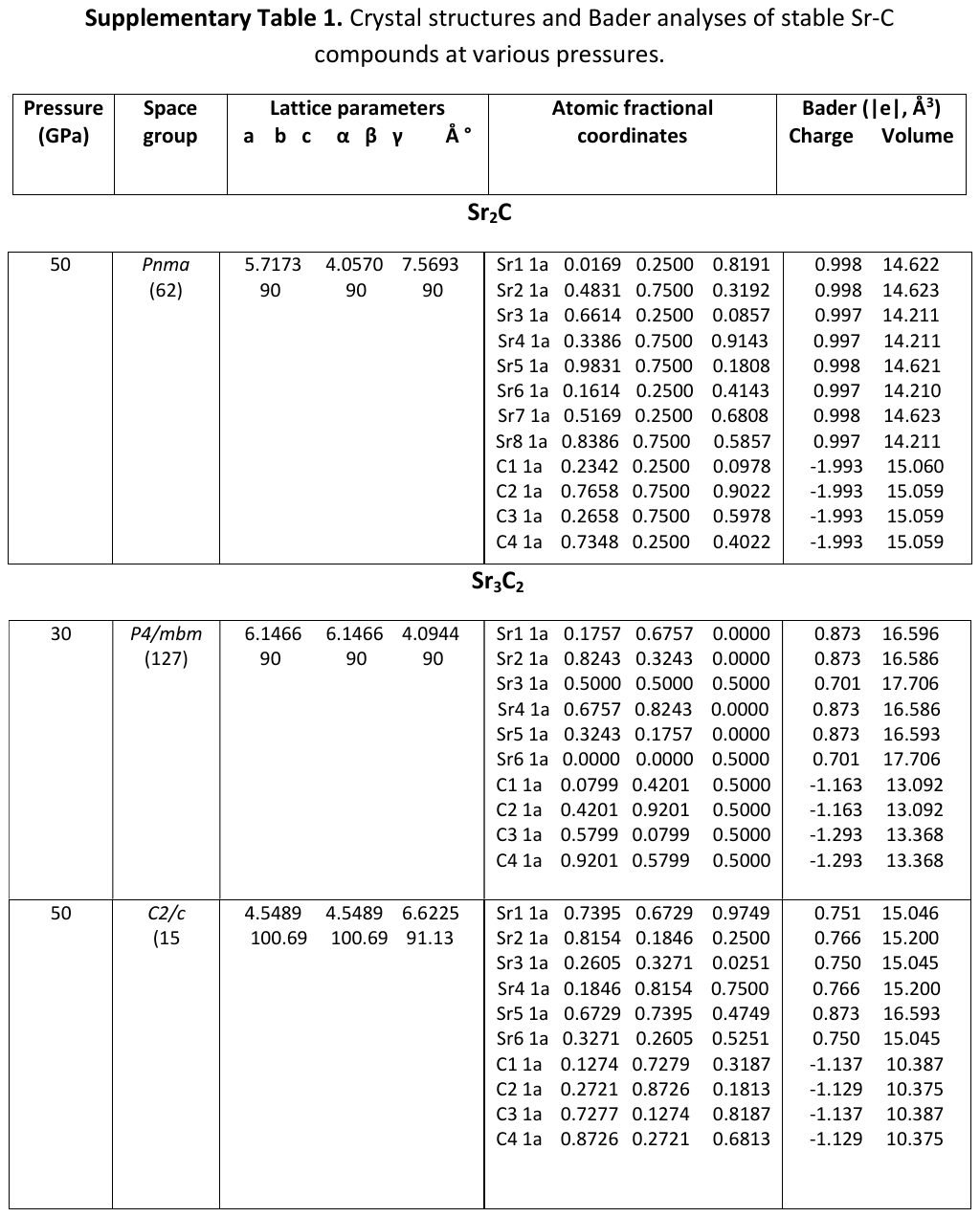}}
      \label{table1:src}
\end{figure*}

\begin{figure*}[hbt!]
	\centering
	\makebox[0pt][c]{\includegraphics[width=1.2\linewidth]{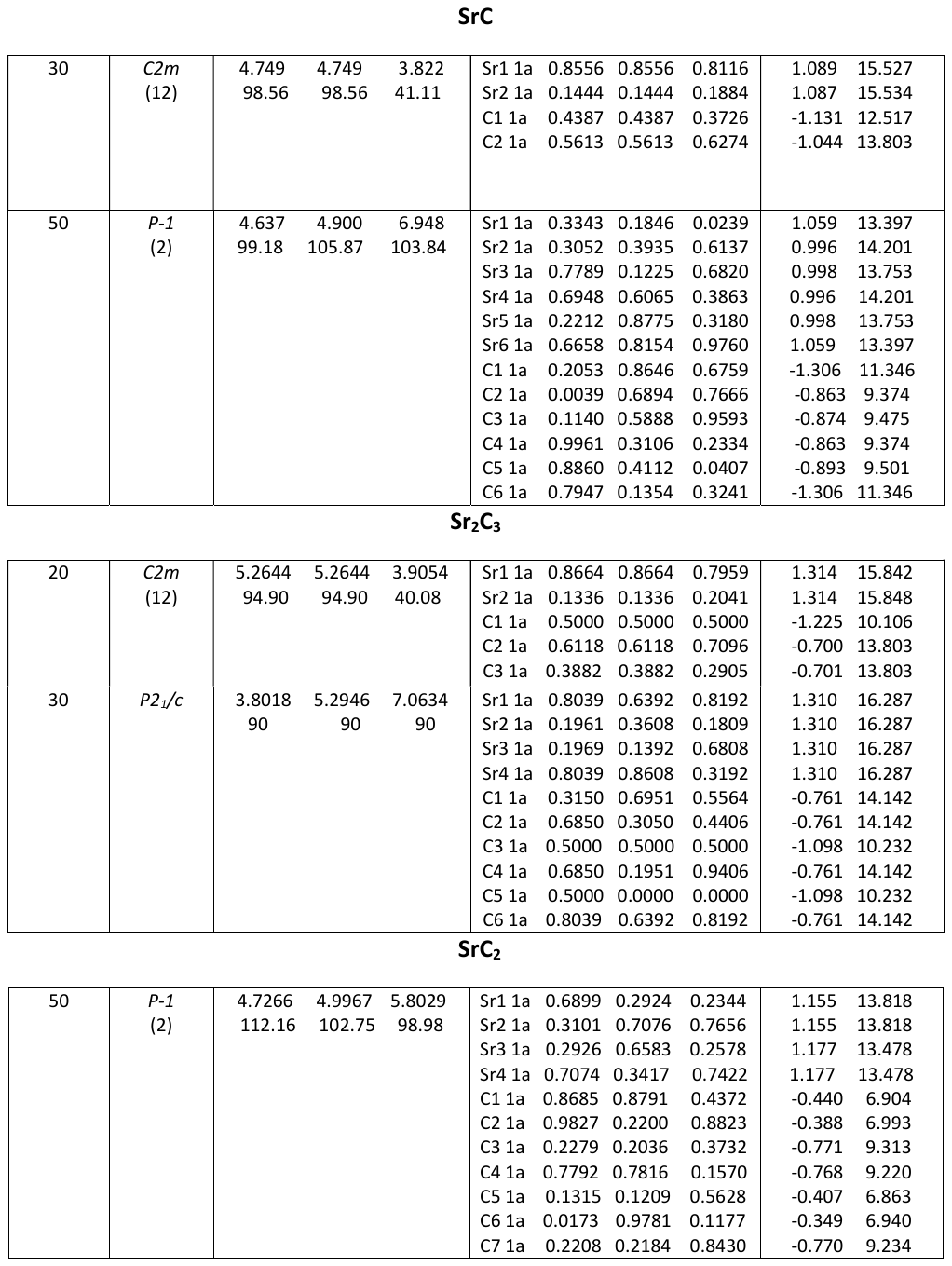}}
      \label{table2:src}
\end{figure*}

\begin{figure*}[hbt!]
	\centering
	\makebox[0pt][c]{\includegraphics[width=1.2\linewidth]{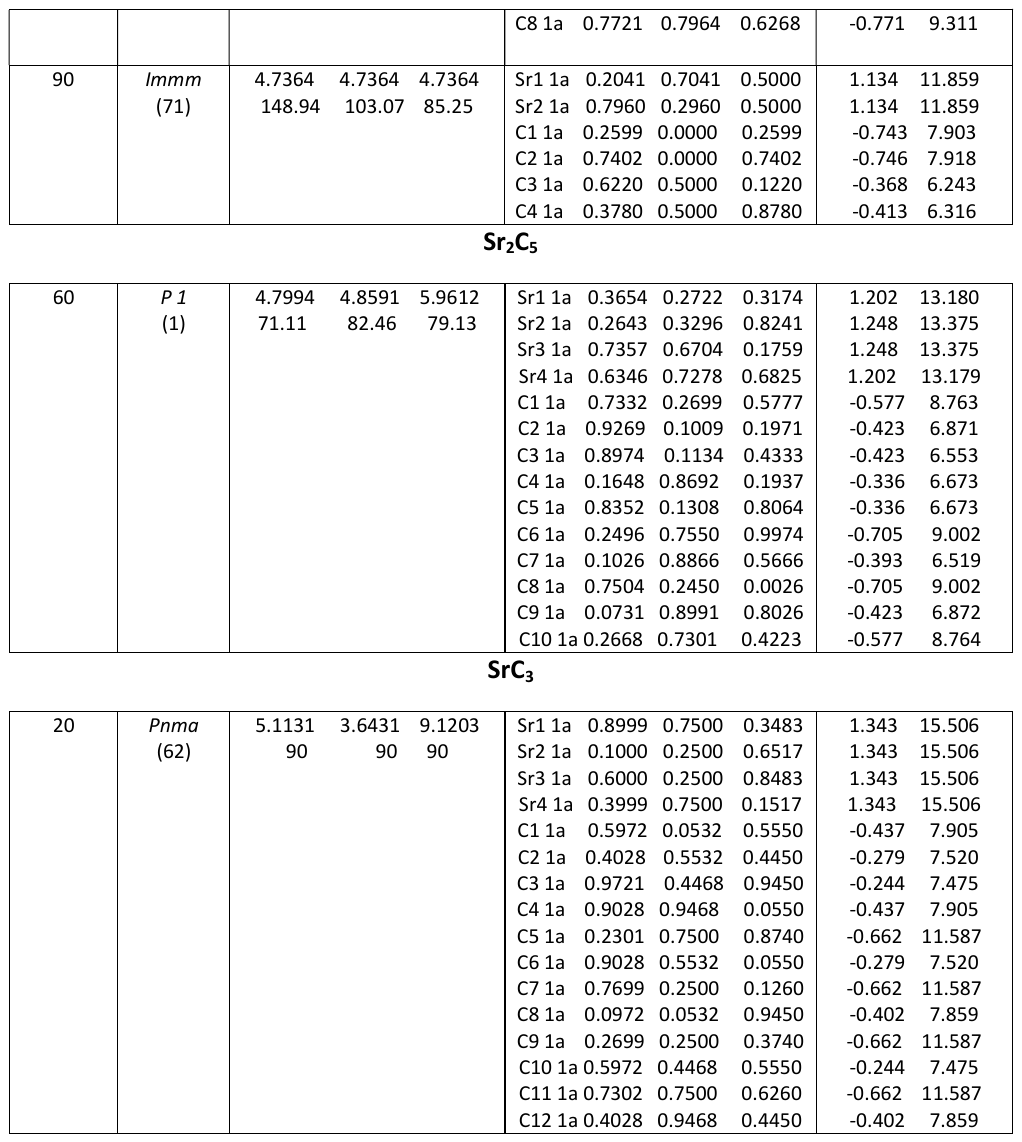}}
\end{figure*}